\documentstyle[prd,aps,floats,epsf]{revtex}

\def\bea{\begin{eqnarray}}
\def\eea{\end{eqnarray}}

\begin{document}
\preprint{BROWN-HET-1213, PURD-TH-00-04, hep-ph/yymmddd}
\draft 

%
%
%
\renewcommand{\topfraction}{0.99}
\renewcommand{\bottomfraction}{0.99}
\twocolumn[\hsize\textwidth\columnwidth\hsize\csname 
@twocolumnfalse\endcsname
  
\title
{\Large Parametric Amplification of Metric Fluctuations During 
Reheating in Two Field Models}
\author{F. Finelli$^{1,2}$ and R. Brandenberger$^3$} 
\address{~\\$^1$Department of Physics, Purdue University, 
West Lafayette,
IN 47907, USA;
~\\$^2$Department of Physics and INFN, I-40126 Bologna, Italy;
~\\$^3$Department of Physics, Brown University,  
Providence, RI 02912, USA;}
\date{\today} 
\maketitle
\begin{abstract} 
We study the parametric amplification of super-Hubble-scale scalar 
metric fluctuations at the end of inflation in some specific two-field 
models of inflation, a class of which is motivated by hybrid inflation. 
We demonstrate that there can 
indeed be a large growth of fluctuations due to parametric resonance 
and that this effect is not taken into account by the 
conventional theory of isocurvature perturbations. Scalar field 
interactions play a crucial role in this analysis. 
We discuss the conditions under which there can be nontrivial parametric 
resonance effects on large scales.        
\end{abstract}

]

\vskip 0.4cm
\section{Introduction}
 
It was recently suggested \cite{BKM1} that parametric 
resonance during the reheating phase of an inflationary Universe 
\cite{TB90,KLS,STB95} 
may lead to an exponential amplification of 
super-Hubble scale gravitational fluctuations. If true, this would affect 
the usual predictions of inflationary models for observables such as the 
matter power spectrum and the spectrum of cosmic microwave anisotropies. 
In particular, it would require the coupling constants in the particle 
physics model of inflation to be exponentially smaller than previously 
thought in order that the theory does not generate a too large amplitude 
for the fluctuations.

In Ref. \cite{FB1} it was shown that, although there are no 
causality constraints which prohibit the amplification of super-Hubble 
(but sub-horizon) modes during reheating, the effect does not occur in a 
simple massive scalar field model of chaotic inflation based on the 
potential $V(\phi) = m^2 \phi^2 / 2$ (Here, $\phi$ is the inflaton field). 
This is true even beyond the linear analysis \cite{PE1}. Similarly, there 
is no effect for a quartic potential \cite{EP2} 
$V(\phi) = \lambda \phi^4 / 4$, nor for a
potential containing both quadratic and quartic terms \cite{LMZ99}. 
These results agree with the earlier analyses in Refs. \cite{NT97} 
and \cite{KH96}.

It was then suggested \cite{BTKM2} that the amplification 
of super-Hubble-scale modes would occur for two field models of inflation, 
e.g. for a model with potential
\begin{equation} \label{KLSmodel}
V(\phi, \chi) \, = \, {1 \over 2} m^2 \phi^2 + 
{1 \over 2} g^2 \phi^2 \chi^2 \, ,
\end{equation}
where as before $\phi$ is the inflaton field and $\chi$ is a second 
scalar matter field. This model had earlier been analyzed by Taruya and 
Nambu \cite{TN98} who claimed that the isocurvature mode of the fluctuations 
will be parametrically amplified during reheating. 
However, as was shown in Refs. \cite{JS} and \cite{Ivanov} 
(and more recently in \cite{LLMW}), the fluctuations in the $\chi$ field are 
exponentially suppressed during 
inflation for values of the coupling 
constant for which the equation of motion of the metric fluctuations 
corresponds to broad resonance, thus rendering the effect studied in 
Ref. \cite{BTKM2} completely inefficient.

The suppression of fluctuations in the $\chi$ field which renders the 
parametric amplification of gravitational fluctuations ineffective in the 
model given by (\ref{KLSmodel}) occurs since during inflation the induced 
mass $m_{\chi}$ of the $\chi$ field which is given by 
$m_{\chi} = g |\phi|$ is larger than the Hubble expansion parameter 
$H$, and hence, as can be easily seen by considering the equation of motion
\begin{equation} \label{chieom}
{\ddot \delta \chi} + 3 H {\dot \delta \chi} 
+ \bigl( {{k^2} \over {a^2}} + g^2 \phi^2 \bigr) \delta \chi \, = \, 0 
\end{equation}
for the linearized fluctuation of the $\chi$ field with comoving 
wave number $k$ (the scale factor is denoted by $a(t)$), 
$\delta \chi$ undergoes damped oscillatory motion.

A model in which $m_{\chi} < H$ during the stage of inflation 
when scales of cosmological interest today exit the Hubble radius 
was recently studied by Bassett and Viniegra \cite{BV99}. 
It is a two field model given by the potential
\begin{equation} \label{GKLSmodel}
V(\phi, \chi) \, = \, {1 \over 4} \lambda \phi^4 
+ {1 \over 2} g^2 \phi^2 \chi^2 \, . 
\end{equation}
In the absence of metric fluctuations, 
this model was studied in detail in 
Ref. \cite{GKLS} (see also \cite{Kaiser}), where it was shown 
that for values of the coupling constants satisfying
\begin{equation}
g^2 \, \simeq \, 2 \lambda
\end{equation}
long wavelength modes ($k \simeq 0$) are in the first 
broad instability band of the Floquet-type equation of motion 
derived from (\ref{chieom}) after field rescaling which describes 
the parametric resonance of matter fluctuations in an unperturbed expanding 
space-time. Bassett and Viniegra \cite{BV99} showed that in this model
the quantity $\zeta$ \cite{BST83} increases exponentially during the
initial stages of reheating. Note that $\zeta$ is a measure of the 
curvature fluctuations and is believed to be conserved on super-Hubble 
scales in the absence of isocurvature fluctuations (see e.g. \cite{MFB92} 
for a review of the theory of cosmological fluctuations). However, since 
the model given by (\ref{GKLSmodel}) admits isocurvature fluctuations, 
a growth of $\zeta$ on super-Hubble modes is expected also in the ``usual" 
analysis of the evolution of fluctuations in inflationary cosmology.

In this paper we take a closer look at the theory given by the potential 
(\ref{GKLSmodel}). Subject to certain assumptions on initial conditions
of the background field dynamics we 
recover results similar to Bassett and Viniegra: exponential growth of
$\zeta$ during the initial stages of reheating. 
Furthermore, we demonstrate that this effect is not taken into account 
by the conventional theory of isocurvature perturbations. 
We then discuss some criteria which an inflationary Universe model must 
satisfy in order to have substantial parametric growth of $\zeta$ during 
reheating via scalar field interactions. We argue that these conditions are 
naturally satisfied in some models of hybrid inflation, and we study a couple 
of concrete examples in which super-Hubble-scale gravitational fluctuations 
grow exponentially during reheating (some other examples where exponential 
growth of super-Hubble-scale modes could occur are given in \cite{BGMK}). 

\vskip 0.4cm
\section{Massless Two Field Model Reconsidered}

In this section we will consider the two-field model with potential 
(\ref{GKLSmodel}) in which Bassett and Viniegra \cite{BV99} showed 
the parametric resonance of super-Hubble scale gravitational modes. 
This model has been studied in detail in \cite{GKLS} in the absence 
of gravitational perturbations (see also \cite{Kaiser}).

As in our previous paper \cite{FB1}, we shall work in longitudinal 
gauge in which the metric including linearized scalar metric fluctuations 
takes on the form
\begin{equation}
ds^2 \, = \, (1 - 2\Phi)dt^2 - a^2(t)(1 + 2 \Psi)dx^idx_i \,
\end{equation}
where the $x^i$ are the comoving spatial coordinates 
and $t$ is physical time. Since for the matter model considered 
the off-diagonal components of the spatial part of the energy-momentum 
tensor vanish, the corresponding components of the Einstein equations 
imply \cite{MFB92} $\Psi = \Phi$. For the moment we will write down the 
equations for a general system with multiple scalar fields ($i = 1, ..., n$), 
and only at a later stage will we specialize to the specific model 
considered in this section. 

The remaining independent equations of motion for linearized perturbations 
in this Einstein-Higgs system are the perturbed energy constraint and 
momentum constraint equations as well as the equations of motion for the 
Higgs field perturbations $\delta \phi_i$:
\begin{eqnarray}
- 3H\dot{\Phi} &-& \left(\frac{k^2}{a^2} + 3 H^2 \right)\Phi \label{adm}\\
&=& 4 \pi G \sum_{i=1}^n \left[\dot{\phi}_i \dot{\delta \phi_i} -
\Phi \dot{\phi_i}^2 + V,_i \delta \phi_i \right]\,, \nonumber
\\
\dot{\Phi} &+& H \Phi =
4 \pi G \sum_{i=1}^n \dot{\phi}_i \delta \phi_i \,,
\label{mom} \\
\ddot{\delta \phi_i} &+& 3 H \dot{\delta \phi_i}
+\left[ \frac{k^2}{a^2} \delta \phi_i + \sum_{j=1}^n V,_{ij} \delta
\phi_j \right] \nonumber \\
&=& 4 \dot{\Phi} \dot{\phi}_i - 2 V,_i \Phi \,,
\label{m2}
\end{eqnarray}
where $k$ denotes the comoving wavenumber, $V,_i$ 
indicates the derivative of $V$ with respect to $\phi_i$, and $G$ 
is Newton's constant.

The Sasaki-Mukhanov \cite{VM88} variables for the $n$ matter fields are 
\begin{equation}
Q_i = \delta \phi_i + \frac{\dot \phi_i}{H} \Phi
\end{equation}
and satisfy the following system of equations:
\begin{equation}
\ddot Q_i + 3 H \dot Q_i + \frac{k^2}{a^2} Q_i +
\sum_{j=1}^{n} [ V_{,i j} - \frac{8 \pi G}{a^3} 
(\frac{a^3}{H} \dot \phi_i \dot \phi_j )^.] Q_j
= 0 \,.
\label{multi}
\end{equation}

We will now specialize to our two field model. 
If the homogeneous part of the second scalar field $\chi$ 
vanishes (which will not be true if parametric resonance is to excite 
gravitational fluctuations - see later), then the inflaton $\phi$ 
during the initial stages of reheating 
(when back-reaction effects are negligible) oscillates as follows \cite{GKLS}
\begin{equation}
\phi(\eta) \, = \, a^{-1} \phi_0 cn(x - x_0, {1 \over {\sqrt{2}}}) \, ,
\end{equation}
where $\phi_0$ is the amplitude of $\phi$ at the end of the slow-rolling 
period, $cn$ is the Jacobi elliptic cosine function, $\eta$ is conformal 
time and $x = {\sqrt{\lambda}} \phi_0$ is a rescaled dimensionless conformal 
time coordinate. Following \cite{GKLS}, it is insightful to rescale 
all the fields $f$ by the scale factor and use new fields ${\tilde f} = a f$. 
The equations of motion (\ref{multi}) then become 
\begin{eqnarray}
{\tilde {Q_{\phi}}}^{''} &+& \bigl[  \kappa ^2 - {{a^{''}} \over a} 
+ 3 cn^2(x - x_0, {1 \over {\sqrt{2}}}) \bigr] {\tilde {Q_{\phi}}} 
\label{phieq}\\
&=& - 2 {{g^2} \over {\lambda \phi_o}} cn(x - x_0, {1 \over {\sqrt{2}}}) 
{\tilde \chi} {\tilde {Q_{\chi}}} + M_{\phi \phi} {\tilde {Q_{\phi}}}
+ M_{\phi \chi} {\tilde {Q_{\chi}}} \nonumber \\
{\tilde {Q_{\chi}}}^{''} &+& \bigl[ \kappa^2 - {{a^{''}} \over a} 
+ {{g^2} \over {\lambda}} cn^2(x - x_0, {1 \over {\sqrt{2}}}) \bigr] 
{\tilde {Q_{\chi}}} \label{chieq}\\
&=& - 2 {{g^2} \over {\lambda \phi_o}} cn(x - x_0, {1 \over {\sqrt{2}}})
{\tilde \chi} {\tilde {Q_{\phi}}} + M_{\phi \chi} {\tilde {Q_{\phi}}}+
M_{\chi \chi} {\tilde {Q_{\chi}}} \nonumber
\end{eqnarray} 
where, following the notation of \cite{BV99} 
(up to a prefactor containing $a$), we have have used the abbreviation
\begin{equation}
M_{\phi_1 \phi_2} \, = \, {{8 \pi G} \over {a^2}} \bigl[
\frac{1}{a H} ({\tilde 
{\phi_1}}^{'} - {{a^{'}} \over a} {\tilde {\phi_1}})  
({\tilde {\phi_2}}^{'}
- {{a^{'}} \over a} {\tilde {\phi_2}}) \bigr]' \, ,
\end{equation}
and where
\begin{equation}
\kappa^2 \, = \, {{k^2} \over {\lambda \phi_0^2}} \, .
\end{equation}
In the model considered, the time-averaged equation of state 
is that of radiation and hence the scale factor is linear in $\eta$ 
and thus $a^{''} = 0$.

Neglecting for a moment the terms on the right hand side of the equations, both 
equations (\ref{phieq}) and (\ref{chieq}) are of the form of Lam\'e equations. 
Lam\'e equations in the context of reheating were first noticed in 
\cite{KLS} and then studied in detail in \cite{Boyan96}, \cite{Kaiser} 
and \cite{GKLS}. The coefficient in front of the $cn^2$ term 
(which in the case of (\ref{chieq}) is ${g^2} /\lambda$) is crucial to the 
resonance structure of the equation. Since we are interested in super-Hubble 
modes, we need to know for which values of the coefficient the mode 
$\kappa = 0$ lies in the first instability band of the equation. 
This is the case for values 
\footnote{Note that this condition and the corresponding conditions 
in all the other examples discussed
in this paper are stable against perturbative 
coupling constant renormalizations, the reason being that we consider 
$g^2 \sim \lambda$ {\it and} $\lambda$ should be constrained by the 
CMB anisotropy results to be a small number $<< 1$. 
Thus, the perturbative correction terms which are of order 
$g^4$ are much smaller than either $g^2$ or $\lambda$.}
\begin{equation}
1 \, < \, {{g^2} \over {\lambda}} \, < \, 3 \, .\label{coeffcond}
\end{equation}
This implies that resonance occurs only in the equation (\ref{chieq}) 
for ${\tilde {Q_{\chi}}}$, and then only for values of ${g^2} / \lambda$ 
in the above range (or in the range corresponding to other instability bands). 
However, if the condition (\ref{coeffcond}) is satisfied, and if the 
$\chi$ field is indeed excited, then parametric amplification of 
${\tilde {Q_{\chi}}}$ is expected, and via the terms on the right hand 
side of (\ref{phieq}), induced exponential growth of ${\tilde {Q_{\phi}}}$ 
should occur.

Note, however, that it is not sufficient to show that an amplification 
of ${\tilde {Q_{\phi}}}$ or ${\tilde {Q_{\chi}}}$ occurs in order to 
demonstrate that parametric resonance during reheating will have a crucial 
effect on the amplitude of gravitational fluctuations. Note that also in
the conventional treatment of fluctuations \cite{MFB92} the amplitude of 
the ${\tilde Q}$ variable grows during the interval in which the equation 
of state of the background changes. As emphasized in \cite{FB1}, a straight
forward way to check if the effect discussed here is a new effect is to 
consider the time evolution of the ``traditional conserved quantity" 
$\zeta$ (\cite{BST83}) which gives a measure of the adiabatic component of 
the metric fluctuations. In the multi-field case, 
$\zeta$ is given by \cite{TN98}
\begin{equation}
\zeta = \frac{H}{\sum_j {\dot \phi_j}^2} \sum_i \dot \phi_i Q_{i} \, .
\label{zetadef}
\end{equation}
In the two field case, the evolution equation for $\zeta$ is \cite{GB1}
\begin{eqnarray}
\dot \zeta &=& - \frac{H}{\dot H} {{\nabla^2} \over {a^2}} 
\Phi \label{zetaeq2} \\
&+& \frac{H}{2} 
\left[ \frac{Q_\phi}{\dot \phi} - \frac{Q_\chi}{\dot \chi} \right]
\frac{d}{d t} \left( \frac{\dot \phi^2 - \dot \chi^2}{\dot \phi^2 + \dot
\chi^2} \right) \,.\nonumber
\end{eqnarray}
In the case of a single scalar field, 
the second term vanishes, but the  equation cannot be applied 
during reheating since ${\dot H} = 0$ at times when ${\dot \phi} = 0$.

In the single field case, the evolution equation 
for $\zeta$ which applies also during reheating 
is (for long-wavelength perturbations for which the spatial gradient 
term can be neglected) \cite{FB1}
\begin{equation}
(1 + w) {\dot \zeta} \, = 0 \, . \label{zetacons}
\end{equation}
This shows that $\zeta$ is conserved unless $w = -1$. 
Unless matter is given by an oscillating scalar 
field (in which case $w = -1$ will occur at the turnaround points 
when ${\dot \phi} = 0$), 
Eq. (\ref{zetacons}) implies that the variable $\zeta$ is conserved 
on scales outside the Hubble radius \cite{BST83}. 
However, reheating corresponds to an oscillating inflaton field, in which
case the conclusion that $\zeta$ is constant may break down, 
as discussed in \cite{FB1}.
Nevertheless, in the specific single field models which 
have been analyzed \cite{KH96,NT97,FB1,PE1,EP2,LMZ99} it was found that 
no net increase of $\zeta$ occurs during the initial stages of reheating, 
and a general proof of the constancy of $\zeta$ in single field models 
was suggested in \cite{LMZ99}.

If matter is described, more realistically, 
in terms of multiple scalar fields (each of which is given by a 
conventional action), then it appears extremely unlikely 
that $w = -1$ will occur at all, since at the points in time 
when ${\dot \phi} = 0$, the other fields will not all also be at rest, 
and thus the net value of $w$ will be greater than $-1$. 
Therefore, the only realistic possibility for growth of $\zeta$ is as a 
consequence of the second term on the right hand side of (\ref{zetaeq2}), 
a term which corresponds to an isocurvature perturbation. 
Inspection of (\ref{zetaeq2}), however, immediately shows that during 
reheating there is the chance of having a very large increase in $\zeta$ 
as a consequence of the zeros in ${\dot \phi}$ which arise periodically 
in time.
This effect is missed if the scalar fields are treated in 
the slow-roll approximation, or if the change in the equation of state 
during reheating is modelled as a monotonic change from a nearly 
de-Sitter equation of state during inflation to a radiative 
equation of state after reheating.

However, to see if there is indeed an exponential growth of 
isocurvature perturbations, it is important to take a closer look at the 
equations. 
In the specific two field model of (\ref{GKLSmodel}), symmetric initial 
conditions for the homogeneous part of 
$\chi$ give $\chi = 0$ and ${\dot \chi} = 0$. In this case, 
it follows from (\ref{zetadef}) that $\zeta$ only depends on 
$Q_{\phi}$, and, since by (\ref{phieq}) there is no parametric amplification 
of super-Hubble modes of $Q_{\phi}$ given that the coupling to $Q_{\chi}$ 
vanishes, that there will therefore be no parametric amplification of 
$\zeta$. The same result can also be seen from (\ref{zetaeq2}) since for 
symmetric initial conditions, the time derivative on the right hand side 
of the equation acts on a constant. The result is confirmed by our 
numerical analysis (see Figure 1).

\begin{figure}
\raisebox{4cm}{$(1+w) \zeta$} \hspace{-0.2cm}
\epsfxsize=2.9 in \epsfbox{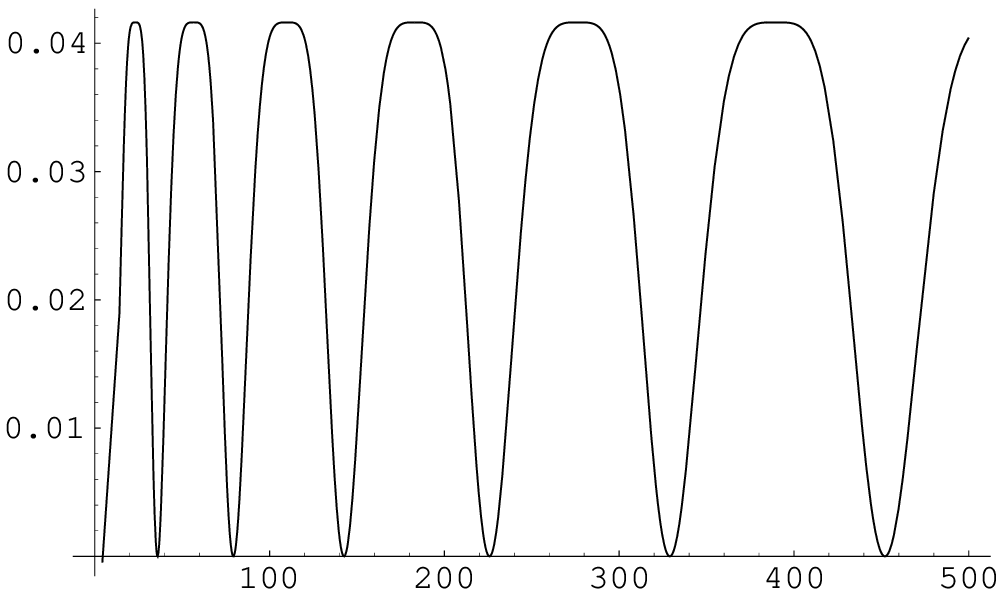} 
\begin{center} $z=\sqrt{\lambda} \, M_{\rm pl} \, t$ \end{center}
\vspace{.2cm}
\caption{Evolution of $(1+w) \zeta$ for the model of Eq. (3) as a function
of the a-dimensional time $z=\sqrt{\lambda} M_{\rm pl} t$ for $\chi_0 =
\dot \chi_0 = 0$ and $\phi_0 = 3.5 M_{\rm pl} \, , \, \dot \phi_0 = - .1 M_{\rm
pl}$ as initial conditions for the background. The fluctuation $Q_\phi$
and $Q_\chi$ start in the adiabatic vacuum $40$ e-foldings before
inflation ends. The wavenumber is $k = 10^2$, which corresponds to five
times the Hubble radius at the beginning of the simulation. Note that the
mode is far outside the Hubble radius at the end of inflation.} 
\label{fig1} 
\end{figure}

However, due to quantum fluctuations we expect that the average of 
$\chi$ over a volume corresponding to a particular super-Hubble 
(but sub-horizon) mode will not vanish. 
It is reasonable \cite{Linde,FV} to use for the homogeneous value 
of $\chi$ the r.m.s. value of the renormalized quantum fluctuations.
It follows from (\ref{chieq}) that $Q_{\chi}$ 
will experience parametric amplification during the initial stages 
of reheating. It will grow as  $exp[\mu_0 \eta]$, where $\mu_0$ is 
the Floquet index of $k=0$ (which
for continuity reasons cannot be too different from the Floquet 
exponent for long wavelengths). 
Via the non-vanishing source terms in (\ref{phieq}), 
this will induce parametric growth of $Q_{\phi}$, 
and this quantity mainly will contribute to the parametric growth of
$\zeta$.
Our numerical analysis confirms the above considerations. 
In Figure 2 we depict the growth of $\zeta$ during the initial stages 
of reheating for two different values of the 
homogeneous component of $\chi$, which shows how the onset of the
parametric growth of $\zeta$ is dependent on the value of $\chi$. In
Figure 3 we show how the growth of $\zeta$ is similar to the growth of
$Q_\phi$.   

\begin{figure}
\raisebox{4cm}{$(1+w)\zeta$} \hspace {-0.2cm}
\epsfxsize=2.9 in \epsfbox{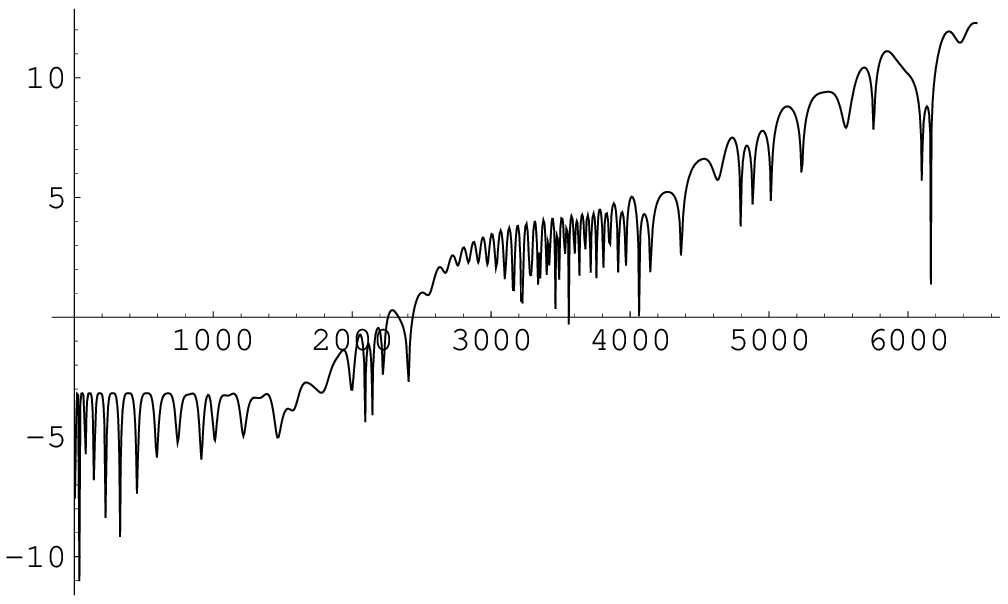} \\
\vspace{.2cm}
\raisebox{4cm}{$(1+w)\zeta$} \hspace {-0.4cm}
\epsfxsize=2.9 in \epsfbox{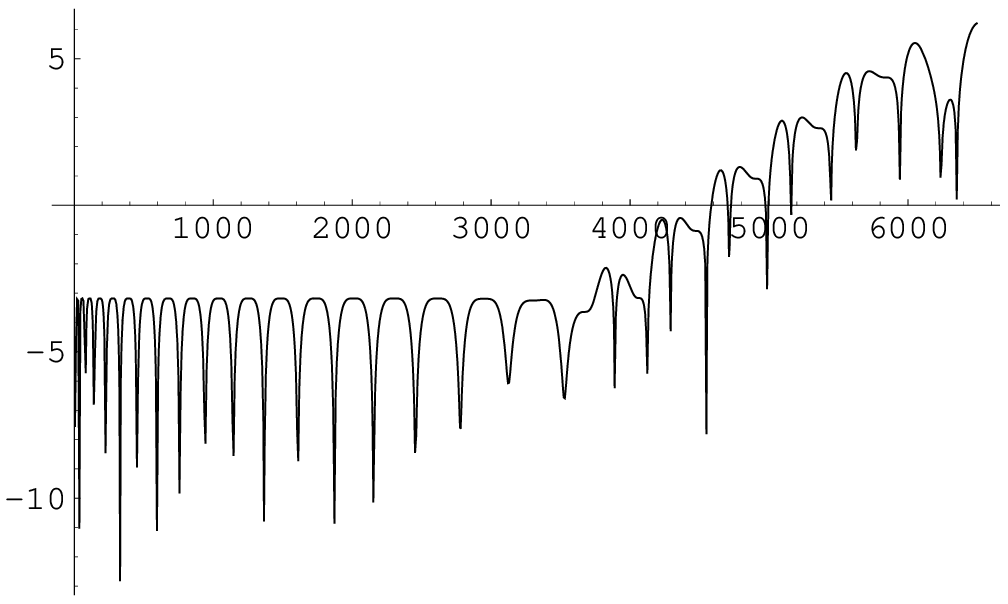}
\begin{center} $z=\sqrt{\lambda} \, M_{\rm pl} \, t$ \end{center}
\vspace{.2cm}
\caption{Evolution in logarithmic scale of $(1 + w) \zeta$ for 
$\chi_0 = 2 \times 10^{-2} M_{\rm pl} \,, \, \dot \chi_0 = 0$ (top) and 
$\chi_0 = 2 \times 10^{-8} M_{\rm pl} \,, \, \dot \chi_0 \sim \lambda
M_{\rm pl} \chi_0$ (bottom) as initial value for
$\chi$. The initial condition for $\phi$ and 
$\dot \phi$ are the same as in Fig. 1 in both of the panels.
The fluctuations $Q_\phi$ and $Q_\chi$ start
in the adiabatic vacuum $40$ e-foldings before
inflation ends. The wavenumber is $k = 10^2$, which corresponds to five
times the Hubble radius at the beginning of the simulation.
The growth of $\zeta$ is delayed in the second case because the background
field, and consequently the mixing terms in Eq. (\ref{phieq}), are smaller
than in the first case: in this way $Q_\chi$ takes longer to feed the
growth of $Q_\phi$ and $\zeta$. The initial conditions for the second
case correspond to the values obtained through renormalization arguments.}
\label{fig2}
\end{figure} 

\begin{figure}
\raisebox{4cm}{$Q_\phi$} \hspace {-0.2cm}
\epsfxsize=2.9 in \epsfbox{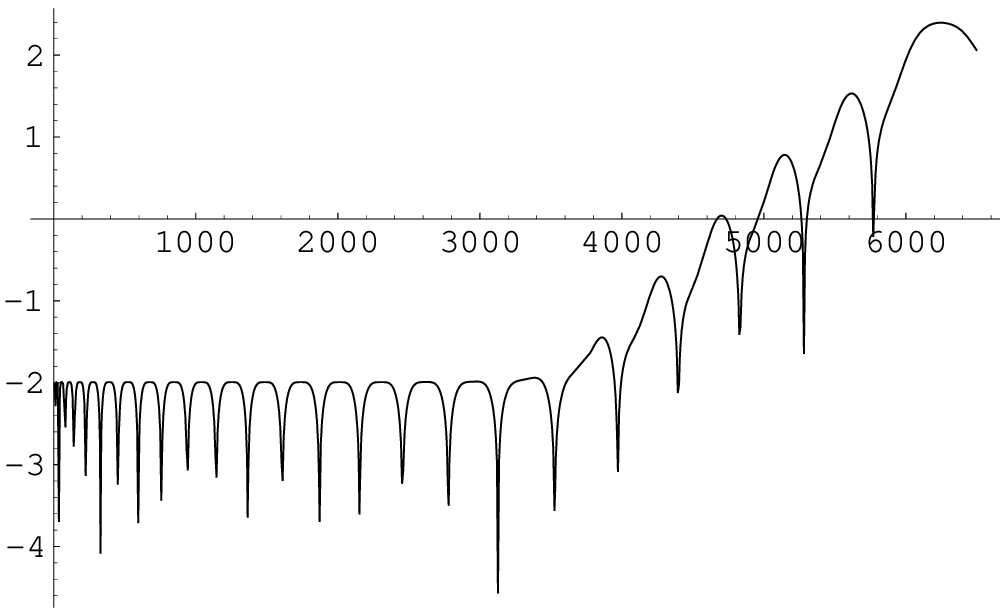} \\
\vspace{.2cm}
\raisebox{4cm}{$Q_\chi$} \hspace {-0.4cm}
\epsfxsize=2.9 in \epsfbox{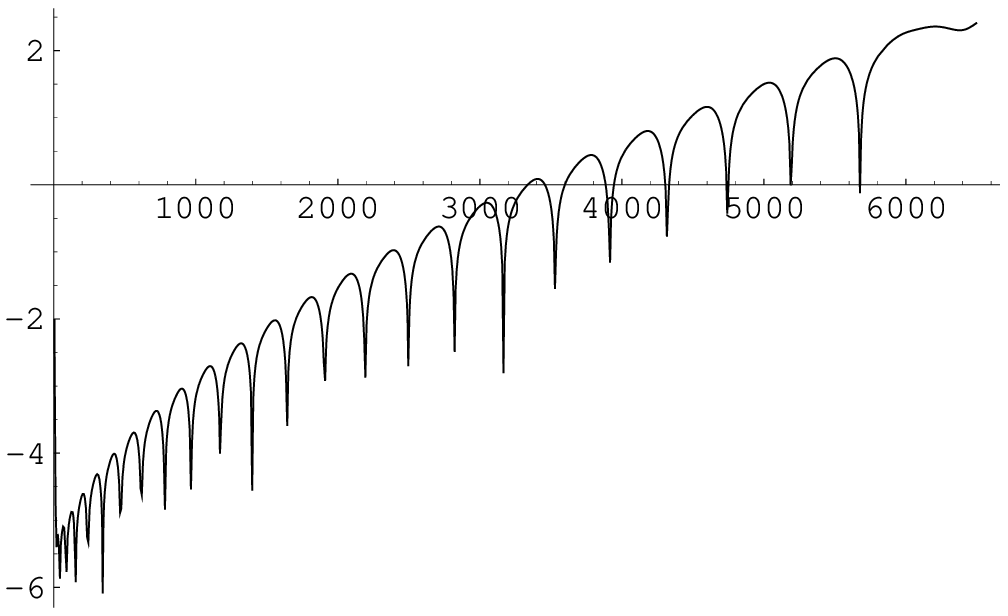}
\begin{center} $z=\sqrt{\lambda} \, M_{\rm pl} \, t$ \end{center}
\vspace{.2cm}
\caption{Evolution in logarithmic scale of $Q_\phi$ (above) and $Q_\chi$
(below) for the second
set of initial conditions of Fig. 2.}
\label{fig3}  
\end{figure}

\vspace{0.4cm}
\section{Isocurvature Perturbations}

Having determined in the previous section that in the model (\ref{GKLSmodel}) 
there is indeed amplification of $\zeta$ during reheating, we must now show 
that this effect is indeed a consequence of parametric resonance, and not 
just an effect due to the change in the equation of state between the 
inflationary era and the post-inflationary era, an effect which is already 
taken care of in the ``usual" theory of isocurvature perturbations in 
inflationary cosmology, which we define as the results obtained when the
transition in the equation of state between the inflationary slow-rolling 
phase ($p \simeq - \rho$) and the post-inflationary radiation-dominated 
phase ($p = {1 \over 3} \rho$) is taken to be monotonic. 
We will show that in the ``usual" analysis there can be no exponential 
increase in the isocurvature perturbation, and that therefore the 
exponential increase we obtain here is a result of parametric resonance. 

The fact that isocurvature perturbations can induce 
an adiabatic component on super-Hubble scales has been known for 
a long time \cite{PV,Bardeen,AW,LM,KS,MFB92}. Entropy perturbations 
act as a source for $\zeta$ even on scales larger than the Hubble radius. 
This is true even in the case when matter is given by a single scalar field. 
In this case, the evolution equation for $\zeta$ becomes \cite{GB2}
\begin{equation}
{\dot \zeta} \, = \, 3 H \bigl( {{\dot p} \over {\dot \rho}} - 
{{\delta p} \over {\delta \rho}} \bigr) \, .
\end{equation}
These perturbations, however, are suppressed on scales larger 
than the Hubble radius.

In models with two or more scalar fields, the equation for $\zeta$ 
is given by (\ref{zetaeq2}), and it is thus clear that even on super-Hubble 
scales one should expect ${\dot \zeta} \neq 0$. In the approximation in which 
both fields are slowly rolling, the time evolution of $\zeta$ on scales 
larger than the Hubble radius was studied in detail in \cite{GB1,GB2,StSa}, 
with particular emphasis on calculating the deviations from scale-invariance 
of the resulting power spectrum of density fluctuations. However, 
since the analyses made use of the slow-rolling approximation, no effects 
of the dynamics of reheating were considered.

More recently, Taruya and Nambu \cite{TN98} and Bassett et al. 
\cite{BTKM2} considered the effect of reheating on the spectrum of 
density fluctuations and discovered a large growth of $\zeta$ due to 
the initial isocurvature perturbations, however in a model in which the 
necessary initial $\chi$ field fluctuations are exponentially suppressed 
during inflation. Bassett and Viniegra \cite{BV99} then pointed out that 
the suppression would be absent in the model (\ref{GKLSmodel}).

It has been known for a long time that isocurvature perturbations can be 
produced in inflationary models with more than one scalar field. 
This issue was initially considered in the context of axion perturbations 
in \cite{ABT83}, extended to more general two field models in \cite{Linde84} 
and studied in detail in \cite{Mollerach}. 
It was discovered that initial super-Hubble-scale isocurvature perturbations 
induce an adiabatic component by the time that the scales re-enter 
the Hubble radius. 

The gauge invariant expression for the total isocurvature perturbation
in a multi-fluid system is \cite{KS}
\begin{equation}
p \Gamma \equiv \sum_i (\delta p_i^{\rm gi} - c_s^2 \delta \rho_i^{\rm
gi}) \,, 
\label{isocurvature}
\end{equation}
where $\delta p_i^{\rm gi}$ and $\delta \rho_i^{\rm
gi}$ are the gauge invariant pressure and density perturbations with
respect to the total matter rest frame and 
the total speed of sound $c_s^2$ is defined as the weighted sum of the 
i-th speed of sound \cite{KS}:
\begin{equation}
c_s^2 \equiv \frac{\dot p}{\dot \rho} =
\frac{1}{\sum_j \dot\phi_j^2} \sum_i c_{s \, i}^2
\dot{\phi}_i^2
\end{equation}
with 
\begin{equation}
c_{s \, i}^2 = 1 + 2 \frac{V,_i}{3 H \dot{\phi}_i} 
\end{equation}
The total isocurvature perturbation can be written as the sum of 
the non-adiabatic pressure component of the single component and of the 
relative isocurvature perturbation $S_{ij}$ as:
\begin{equation}
p \Gamma = \sum_i (\delta p_i - c_{s \, i}^2 \delta \rho_i) +
\frac{1}{\sum_i \dot \phi_i^2} \sum_{i,j} \frac{\dot \phi_i^2 \dot
\phi_j^2}{2} S_{i j} (c_{s\, i}^2 - c_{s \,
j}^2) 
\end{equation}
where
\begin{equation}
S_{ij} \equiv 
\frac{\delta \rho_i^{\rm gi}}{\dot \phi_i^2} - 
\frac{\delta \rho_j^{\rm gi}}{\dot \phi_j^2} \,.
\label{iso}
\end{equation}
The relative isocurvature perturbation $S_{ij}$ with respect to the
total matter frame can be written for our two field model as 
\begin{eqnarray}
S_{\phi \chi} &=&
\frac{\delta \rho_\phi}{\dot \phi^2} - \frac{\delta
\rho_\chi}{\dot \chi^2} - \frac{a v}{k} Q_\phi \left(
\frac{1}{\dot{\phi}^2} + \frac{1}{\dot{\chi}^2} \right)
\nonumber \\
&=& \frac{\delta \rho_\phi}{\dot \phi^2} - \frac{\delta
\rho_\chi}{\dot \chi^2} - \frac{ 3 g^2}{8 \pi G} {{\phi \chi} \over
{{\dot \phi}^2{\dot \chi}^2}}
(\dot \phi \chi - \dot \chi \phi) \frac{\dot H}{H}(\zeta - \Phi)
\,, \label{relent}
\end{eqnarray}
where $v$ is the total perturbed velocity for matter, $Q_\phi$ is the
homogeneous energy transfer to the $\phi$ component ($Q_\phi + Q_\chi = 0$
because of total energy conservation), 
and now $\delta \rho_i$ are the density perturbations in the
longitudinal gauge:
\begin{equation}
\delta \rho_i = \dot \phi_i \delta \dot \phi_i - \Phi \dot \phi_i^2
+ V,_i \delta \phi_i \, .
\end{equation}

{F}rom Equation (\ref{relent}) it follows that the parametric 
resonance from the matter sector of the theory 
(the $Q$ variables to be specific) induces exponential 
growth of the relative isocurvature perturbation, and hence also of 
the total isocurvature perturbation (see Figure 4). 

\begin{figure}
\raisebox{4cm}{$p \Gamma$} \hspace {-0.2cm}
\epsfxsize=2.9 in \epsfbox{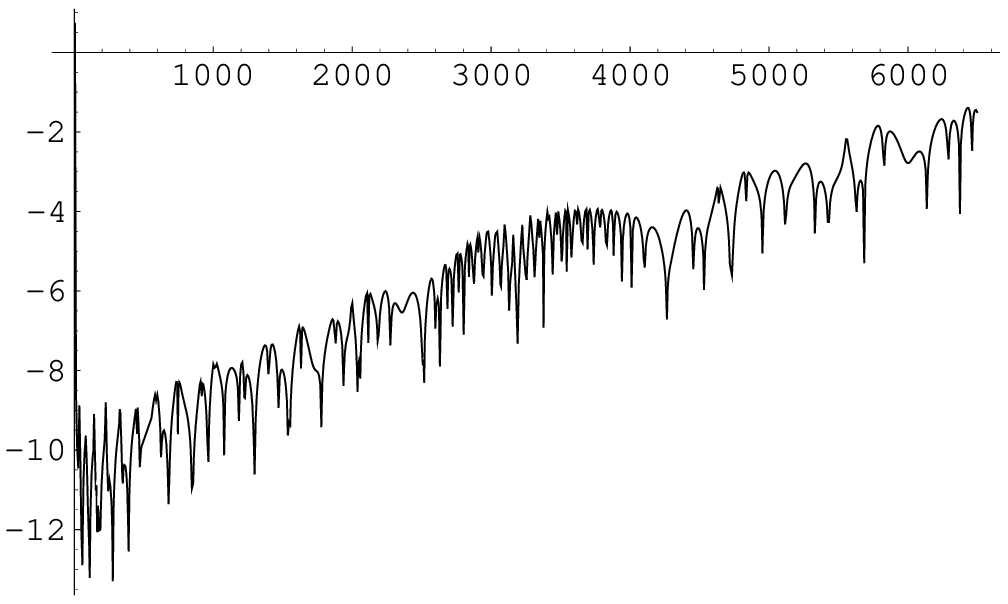} 
\begin{center} $z=\sqrt{\lambda} \, M_{\rm pl} \, t$ \end{center}
\vspace{.2cm}
\caption{Evolution in logarithmic scale of the total non-adiabatic
pressure $p \Gamma$ for the first set of initial conditions of Fig. 2.}
\label{fig4}
\end{figure}

In turn, isocurvature perturbations determine the change in $\zeta$ via 
Eq. (\ref{zetaeq2}).
This shows that in the presence of scalar field interaction terms, there
is a correlated exponential growth of $\zeta$ and of the relative 
isocurvature perturbation $S_{\phi \chi}$. This exponential growth is a 
consequence of parametric resonance and is absent if the phase transition 
is modelled with a monotonically increasing value of $w$.

Analogously to Eq. (\ref{zetadef}) for the Bardeen parameter, the total
non adiabatic pressure $p \Gamma$ can be expressed in terms of the
Sasaki-Mukhanov variables in the following way:
\begin{equation}
p \Gamma = \sum_i \left[ \frac{\dot V}{\rho} \dot{\phi}_i Q_i - 2 V,_i Q_i 
- 2 \frac{\dot V}{3 H \sum_i \dot{\phi}_i^2} (\dot{\phi}_i \dot{Q}_i +
V,_i Q_i) \right] \,.
\label{isodef}
\end{equation} 

As mentioned above, the exponential growth of $S_{\phi \chi}$ and $\zeta$ 
during reheating is a new effect due entirely to parametric resonance. 
The growth of fluctuations in inflationary models with two uncoupled
fields were studied in \cite{Mollerach} in an approximation in 
which the oscillations of the inflaton field were neglected. 
In this case there is no growth of $S_{\phi \chi}$. 
An initial isocurvature perturbation does induce the growth of an 
adiabatic component on super-Hubble scales, but the final amplitude of 
the adiabatic mode is not much larger than 
the initial amplitude of the isocurvature perturbation, in agreement 
with the earlier analysis in \cite{ABT83}.

\section{Three Golden Rules}

Based on the analysis of Section 2, it appears that several conditions are
required in order to have efficient parametric resonance of
super-Hubble-scale metric fluctuations.

1. In the absence of gravitational perturbations there must be
broad-band parametric resonance in the matter sector of the theory
corresponding to isocurvature fluctuations, and $k=0$ must be part
of the resonance band.

2. The fluctuations in the matter field which undergoes parametric
resonance must be effectively massless during inflation.
More precisely, there should be no large net
suppression of these fluctuations before the phase of parametric
resonance.

3. The homogeneous value of the matter field which undergoes resonance
must be non-vanishing. This is the weakest of the three conditions since
it is only required if we work strictly to first order in perturbations. 

We show how these three rules are satisfied in another model with massless
fields, but now based on {\em negative coupling instability} \cite{GPR}. The
potential is the following:
\begin{equation}
V(\phi, \chi) \, = \, {1 \over 4} \lambda \phi^4 - 
{1 \over 2} g^2 \phi^2 \chi^2 + {1 \over 4} \lambda_\chi \chi^4
\label{GPRmodel}
\end{equation}
with the parameter $r \equiv \lambda \lambda_\chi / g^4 > 1$ in order
to have a potential bounded from below \cite{GPR}. 
This model has an attractor for $\chi$ in the point which minimize the 
potential for $\chi$ \cite{GPR}
\begin{equation}
\bar \chi (t) \sim \frac{g}{\sqrt{\lambda_\chi}} \phi (t) \,. 
\label{attractor}
\end{equation}
In this way even the third and weakest of the above conditions is satisfied. 
The second one is easily satisfied because of the negative effective mass
for the $\chi$ fluctuations when the background $\chi$ is small. In order to
find the unstable bands one can use the attractor solution and estimate
the frequency of the inflaton $\phi$ and
of the fluctuations $\delta \chi$ during the period of coherent oscillations:
\begin{eqnarray}
\omega^2_\phi &=& \lambda \phi^2 - g^2 \chi^2 \sim \lambda \phi^2 (1 -
\frac{1}{r}) \equiv \tilde{\lambda} \, \phi^2
\nonumber \\
\omega^2_{\delta \chi} &=& \frac{k^2}{a^2} + 3 \lambda_\chi \chi^2 - g^2
\phi^2 \sim \frac{k^2}{a^2} + 2 g^2 \phi^2 \nonumber \,.
\end{eqnarray}
If $\chi$ is small compared to the inflaton $\phi$, then an unstable band
for $k = 0$ should be located at $2 g^2 = 2 \tilde{\lambda}$. 
This gives a second order equation for $g^2$ whose positive root is:
\begin{equation}
g^2 = \lambda_\chi \frac{ -1 + \sqrt{1 + 4\lambda/\lambda_\chi}}{2} 
\end{equation}

\begin{figure}
\raisebox{4cm}{$(1 + w) \zeta$} \hspace {-0.2cm}
\epsfxsize=2.9 in \epsfbox{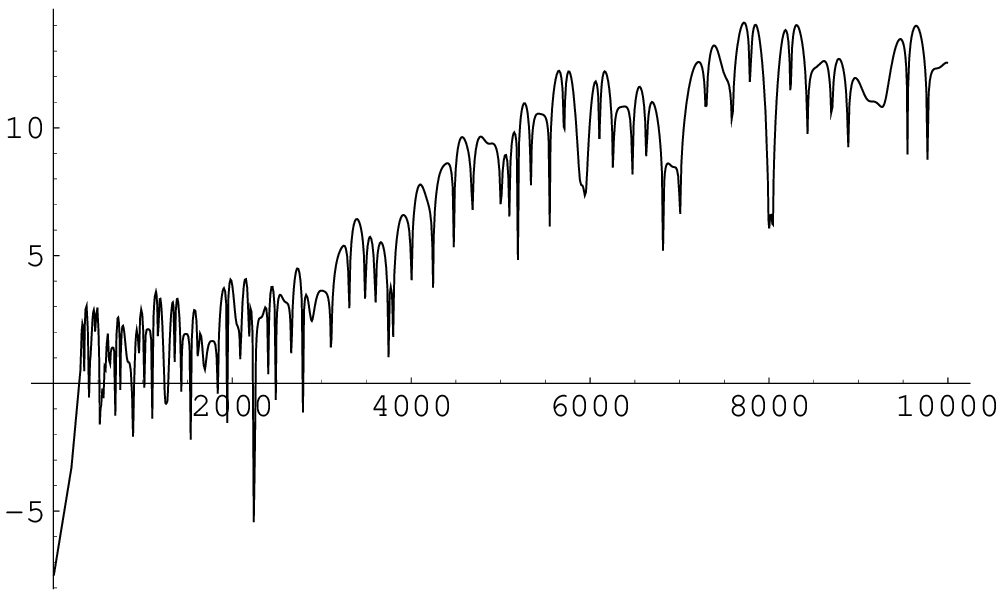} \\
\vspace{.3cm}
\raisebox{4cm}{$(1 + w) \zeta$} \hspace {-0.4cm}
\epsfxsize=2.9 in \epsfbox{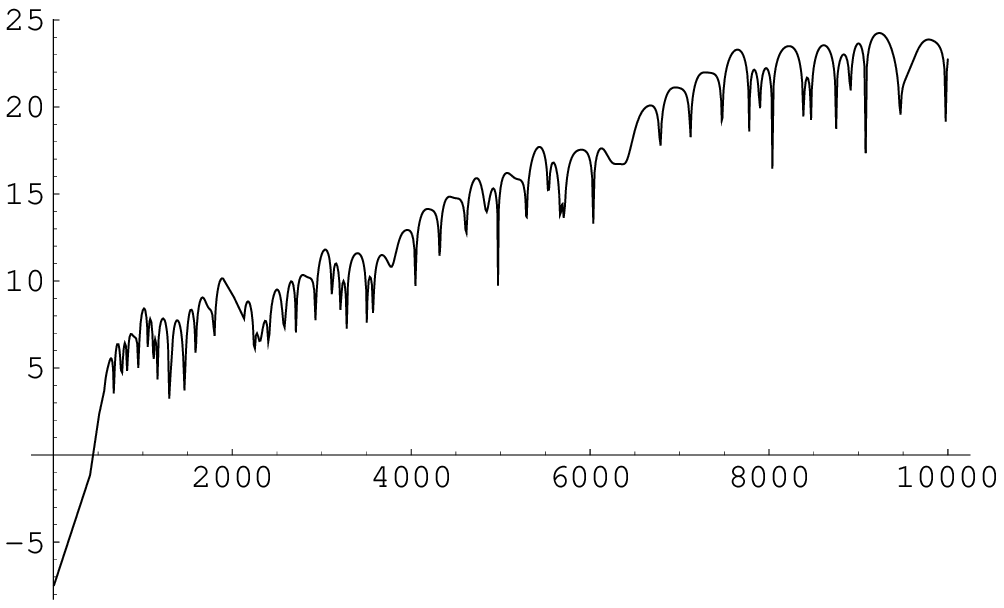}
\begin{center} $z=\sqrt{\lambda} \, M_{\rm pl} \, t$ \end{center}
\vspace{.2cm}
\caption{Evolution in logarithmic scale of $(1 + w) \zeta$ for the model
(\ref{GPRmodel}) for $g^2 = \lambda$, $\lambda_\chi=10^2 \lambda$ (top)
and $\lambda = \lambda_\chi \,, g^2 = \lambda/2 (\sqrt{5} - 1)$
(bottom). The initial conditions for the background are
$\phi_0 = 3.5 M_{\rm pl} \,, \, \dot \phi_0 = -.1 \sqrt{\lambda} M_{\rm
pl} \,, \, \chi_0 = 2 10^{-8} \times M_{\rm pl}$ and $\dot \chi_0 =0$.
The fluctuations $Q_\phi$ and $Q_\chi$ start
in the adiabatic vacuum $40$ e-foldings before
inflation ends. The wavenumber is $k = 10^2$, which roughly corresponds to
five times the Hubble radius at the beginning of the simulation.}
\label{fig5}
\end{figure}

We confirm numerically this analytical estimate in Figure 5 for two
allowed values of $g^2$: $g^2 = \lambda$ with $\lambda \ll \lambda_\chi$
and $g^2 = \lambda/2 (\sqrt{5} - 1)$ with $\lambda = \lambda_\chi$.
The reason for the growth of $\zeta$ is similar to the previous case:
$Q_\chi$ is parametrically amplified, feeds the growth of $Q_\phi$ and
in this case both contribute to the growth of $\zeta$.

\section{A Model Motivated by Hybrid Inflation}

Another natural scenario in which the above conditions can all be
satisfied is
hybrid inflation \cite{hybrid}. 
Hybrid inflation is also an attractive framework for implementing 
inflation in the context of 
supergravity models \cite{hybridsugra}.
Since (at least) two fields are involved in the dynamics of hybrid inflation, 
the generation of isocurvature perturbations is rather natural. 
In hybrid models, the phase of inflation during which the inflaton 
field $\phi$ is slowly rolling towards $\phi = 0$ is terminated by a phase
transition in the second scalar field $\chi$, a field with the double-well 
potential. This implies that during the oscillations of $\phi$, 
the background value of $\chi$ is non-vanishing, leading to an obvious 
realization of condition (3) above.

Parametric resonance in the matter sector of hybrid inflation models 
was studied in detail by Garcia-Bellido and Linde \cite{GBL}, and, in 
supersymmetric hybrid inflation, by Bastero-Gil et al. \cite{BGKS}. The 
resonance of the fluctuations of the two fields $\phi$ and $\chi$ is 
inefficient for a large set of the parameter space since both the $\delta
\phi$ and $\delta \chi$ fields are 
effectively massive during the regime of coherent oscillations 
($\delta \chi$ becomes massive through the Higgs mechanism). Quite
generically, parametric resonance could be much more efficient if there 
is a third field $\psi$ which couples to both $\phi$ and $\chi$.  
Another interesting possibility is to consider a doublet for the field
$\chi$. Then, even in the phase of coherent oscillations, there is a
massless degree of freedom, namely the ``Goldstone" mode. Such a situation
arises naturally in supergravity models \cite{hybridsugra}.

Therefore, as a toy model we will consider the following potential for the 
inflaton field $\phi$ and the doublet 
$\chi = {1 \over {\sqrt{2}}}(\chi_1, \chi_2)$:
\begin{eqnarray}
V \, &=& \, \lambda ({{M^2} \over {2 \lambda}} - |\chi|^2)^2 + {1 \over 2}
m^2 \phi^2 + g^2 \phi^2 |\chi|^2 \nonumber \\
&=& {1 \over 2} m^2 \phi^2 + {1 \over 2} g^2 \phi^2 (\chi_1^2 + \chi_2^2) 
+ {{M^4} \over {4 \lambda}} \nonumber \\
&-& {1 \over 2} M^2 (\chi_1^2 + \chi_2^2) + {{\lambda} \over 2} 
\chi_1^2 \chi_2^2  \nonumber \\
&+& {{\lambda} \over 4}(\chi_1^4 + \chi_2^4) \, .
\end{eqnarray} 
For supersymmetric hybrid inflation, 
there is only one independent coupling constant since 
\begin{equation}
\lambda \, = \, {{g^2} \over 2} \, .
\end{equation}

The values of the masses and coupling constants are 
constrained by the amplitude of density fluctuations at Hubble 
radius crossing, which is given by \cite{hybrid}
\begin{equation} \label{pars}
{{\delta \rho} \over {\rho}} \, \sim \, 
{g \over {\lambda^{3/2}}} \bigl({M \over {M_{pl}}}\bigr)^3 
\bigl( {M \over m}\bigr)^2 \,
\end{equation}
which should be about $10^{-5}$. In our numerical simulations, 
we choose $g^2 = 10^{-3}$, $\lambda = g^2/2$ (as in a supersymmetric 
model), $M^2 / M_{pl}^2 = 10^{-12}$ and $M^2 / m^2 = 10^{10}$. With these 
values, the Hubble parameter during the late stages of inflation is much 
larger than $m$ which ensures slow rolling of $\phi$.

In this model, slow-roll inflation takes place while the value of the 
inflaton $\phi$ is larger than $\phi_c = M/g$. For these values, the 
effective square mass $m_{\chi}^2$ of $\chi$ (evaluated at $\chi = 0$) 
is positive. Once $\phi$ drops below $\phi_c$, the $m_{\chi}^2$ turns 
negative and inflation ends (at time $t_c$) via a symmetry breaking 
transition in the matter sector. We choose the basis of the $\chi$ fields 
such that the order parameter of the transition is $\chi_1$. However, since 
in general the initial ratio of $\chi$ fields and $\chi$ velocities at $t_c$ 
is not the same
\begin{equation}
{{\dot \chi_2} \over {\dot \chi_1}}(t_c) \, \neq \, 
{{\chi_2} \over {\chi_1}}(t_c) \, ,
\end{equation}
we can with no loss of generality assume that the value of $\chi_2$ 
at the time $t_r$, when the $\chi$ transition is complete and the $\phi$ 
oscillations start, does not vanish (as in the previous case 
a reasonable value for $\chi_2$ is the r.m..s. of the renormalized quantum 
fluctuations). Thus, we
have argued that the third 
of the conditions mentioned at the beginning of this section 
(non-vanishing background matter fields) is naturally satisfied in 
this model (in contrast to the model considered in Section 2). 
The initial values of the matter fields at the beginning of the period of 
oscillation will be
\begin{eqnarray}
\chi_1(t_r) &\simeq& M/ {\sqrt{\lambda}} \nonumber \\
0 \neq \chi_2(t_r) &\ll& M/{\sqrt{\lambda}} \, .
\end{eqnarray}

In contrast to the model (\ref{KLSmodel}) 
in which the matter fluctuations are exponentially suppressed 
during inflation, the effective negative coupling instability in the 
matter fields in the time interval between $t_c$ and $t_r$ leads to the 
conclusion that the $\chi$ fluctuations are not suppressed. 
In fact, they are suppressed during slow-roll inflation, but build up
again exponentially fast in the time interval $t_c < t < t_r$. 
To see this, we focus on the evolution of the fluctuations in of the 
$\chi$ field since, as we shall see below, 
these are essential for the effectiveness of parametric resonance. 
We will consider field fluctuations neglecting metric perturbations and 
the mixing terms deriving from particle interactions (even if in the
hybrid models these mixing terms are not perturbatively small and their
importance has been emphasized in \cite{BGKS}).
Under these approximations the evolution equation for $\delta \chi_2$ is
\begin{equation} \label{chi2eq}
{\ddot \delta \chi_2} + 3 H {\dot \delta \chi_2} \, = \, - 
\bigl( {{k^2} \over {a^2}} + g^2 \phi^2 + \lambda \chi_1^2 + 
3 \lambda \chi_2^2 - M^2 \bigr) \delta \chi_2
\end{equation}
For $\phi > \phi_c$, the effective square mass is larger 
than $H^2$ and positive, thus leading to damped oscillatory solutions
\begin{equation}
\delta \chi_2 \, \sim \, a^{-3/2}(t) exp(i \omega t) \, ,
\end{equation}
with $\omega = g \phi$ (in the adiabatic limit).
During this time interval, however, the homogeneous components of 
the matter fields are also damped. The evolution equation for the order 
parameter $\chi_1$ is
\begin{equation} \label{chi1eq}
{\ddot \chi_1} + 3 H {\dot \chi_1} \, 
= \, - \bigl( g^2 \phi^2 + \lambda \chi_1^2 + 
\lambda \chi_2^2 - M^2 \bigr) \chi_1
\end{equation}
Since (up to the contributions from $\chi_2^2$ which are negligible 
during inflation) the effective 
masses in (\ref{chi1eq}) and (\ref{chi2eq}) are the same, the damping 
rates of $\chi_1$ and $\delta \chi_2$ are also the same for $t < t_c$. 
In the time interval between $t_c$ and $t_r$, the signs of the effective
square masses in both Equation (\ref{chi1eq}) for $\chi_1$ and Equation 
(\ref{chi2eq}) for $\delta \chi_2$ are reversed. In both cases, the 
effective $m^2$ is now $- M^2$, leading to exponential increase 
in both $\chi_1$ and $\delta \chi_2$. This period ends
when $\chi_1$ reaches the minimum of the potential at time $t_r$. 
To summarize the above discussion, the evolution of $\chi_1$ follows
\begin{eqnarray} \label{chi1eq2}
{M \over {\sqrt{\lambda}}} \simeq \chi_1(t_r) &\sim&  e^{M(t_r - t_c)} 
\chi_1(t_c) \\
&\sim&  e^{M(t_r - t_c)} e^{-{3 \over 2}H(t_c - t_i)} \chi_1(t_i) \, \nonumber 
\end{eqnarray}
where $t_i$ is the time at the beginning of inflation 
and $H$ is the Hubble constant during inflation, assumed 
to be constant to make the equation simple (this assumption 
does not affect the basic point we are making).
In comparison, the evolution of $\delta \chi_2$ obeys
\begin{eqnarray} \label{chi2eq2}
\delta \chi_2(t_r) &\sim&  e^{M(t_r - t_c)} \delta \chi_2(t_c) \\
&\sim&  e^{M(t_r - t_c)} e^{-{3 \over 2}H(t_c - t_i)} 
\delta \chi_2(t_i) \, \nonumber
\end{eqnarray}
which shows that the exponential growth of $\delta \chi_2$ 
for $t_c < t < t_r$ precisely makes up for the exponential decay during 
the period $t_1 < t < t_c$,
exactly as it does for the evolution of $\chi_1$. Equations (\ref{chi1eq2}) 
and (\ref{chi2eq2}) can be combined to give
\begin{equation}
\delta \chi_2(t_r) \, \sim \, {{M / {\sqrt{\lambda}}} 
\over {\chi_1(t_i)}} \delta \chi_2(t_i) \, .
\end{equation}
This demonstrates that there is no overall suppression of the fluctuations 
in $\delta \chi_2$ before the onset of parametric resonance, showing that 
the second condition for the effectiveness of parametric amplification of 
super-Hubble gravitational modes mentioned at the beginning of this
section is satisfied.

The final conditions to discuss are the criteria for parametric 
resonance of the $k = 0$ modes of the matter perturbations. 
To do this, we consider the mode equation for $\delta \chi_2$ during the 
period in which the inflaton $\phi$ oscillates. For general hybrid models, 
the background dynamics is chaotic since both $\phi$ and $\chi_1$ oscillate 
with different frequencies. However, in the supersymmetric case \cite{BGKS}, 
the frequencies coincide and the background dynamics becomes non-chaotic. 
Both $\phi$ and $\chi_1$ oscillate with the frequency $\sqrt{2} M$. To
simplify the analysis, we shall neglect the back-reaction of
particle production and expansion on the inflaton, and neglect the Hubble 
damping term in the equation of motion (this is a good approximation
since we are considering a case in which $H << \sqrt{2} M$ and the
fields oscillation are not damped by the expansion of the universe).
Therefore, we take the inflaton 
to oscillate with amplitude $\phi_a < \phi_c$, and $\chi_1$ will oscillate 
about its ground state as
\begin{equation}
\chi_1(z) \, = \, {M \over {\sqrt{\lambda}}} (1+ f(z)) \, ,
\end{equation}
where $f(z)$ is periodic with period $2 \pi$. It is convenient to introduce 
the dimensionless time $z = \sqrt{2} M t$. Denoting the derivative with 
respect to $z$ by a prime, the equation for the Fourier mode $\chi_{2 k}$ 
of $\delta \chi_2$ becomes
\begin{equation} \label{res}
\chi_{2 k}^{''} + \chi_{2 k} \bigl( {{k^2} \over {2 a^2 M^2}} 
+ {{g^2 \phi_a^2} \over {4 M^2}} + {{g^2 \phi_a^2} \over {4 M^2}} cos(2z) 
+ f + \frac{f^2}{2}\bigr) \, = \, 0 \, ,
\end{equation}
where we have neglected the terms in $\chi_2$.
In the absence of the final term (the term containing $f(z)$), 
this has the form of the Mathieu equation
\begin{equation}
\chi_{2 k}^{''} + \chi_{2 k} \bigl( A(k) - 2q cos(2z) \bigr) \, = \, 0 \, .
\end{equation}
The value of $q$ is $q \leq 1/8$, the maximal value being taken 
on if $\phi_a = \phi_c$, and for long wavelengths $A(k) \simeq 2q$. 
As can be seen from the
Floquet instability charts (see e.g. Fig. 1 in \cite{GPR}), 
these values do not correspond to efficient resonance. 
{F}rom the evolution of the background fields obtained from the full numerical 
solution of the background field equations (see Figure 6) it follows that 
the amplitude of oscillation $\phi_a$ is in fact substantially smaller than 
$\phi_c$. In contrast, $\chi_1$ oscillates with a large amplitude. Hence, 
the term containing $f(z)$ in Equation (\ref{res}) is more important. 
This term leads to negative coupling instability 
(see \cite{GPR} for a general discussion of resonant particle production 
by negative coupling instability) for small values of $k$. 
Hence, we expect parametric amplification of long wavelength 
gravitational fluctuations in our model.
 
The above considerations are supported by our numerical results. 
In Figure 6 we show the evolution of the background fields $\phi$, 
$\chi_1$, $\chi_2$ and $H$ as a function of time in a simulation 
with parameters mentioned after Equation (\ref{pars}), and
with initial conditions $\phi_0 = 3 M_{\rm{pl}}$, $\chi_1 = .01
M_{\rm{pl}}$, $\chi_2 = .0001 M_{\rm{pl}}$. 

\begin{figure}
\raisebox{4cm}{$\phi$} \hspace {-0.2cm}
\epsfxsize=2.9 in \epsfbox{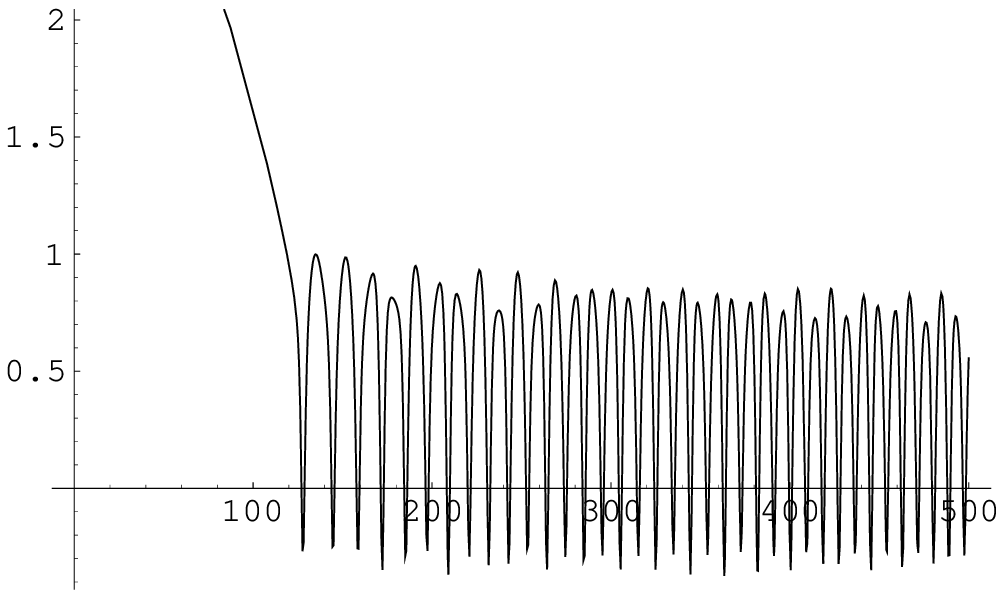} 
\\
\vspace{.3cm}
\raisebox{4.2cm}{$\chi_1$} \hspace {-0.5cm}
\epsfxsize=2.9 in \epsfbox{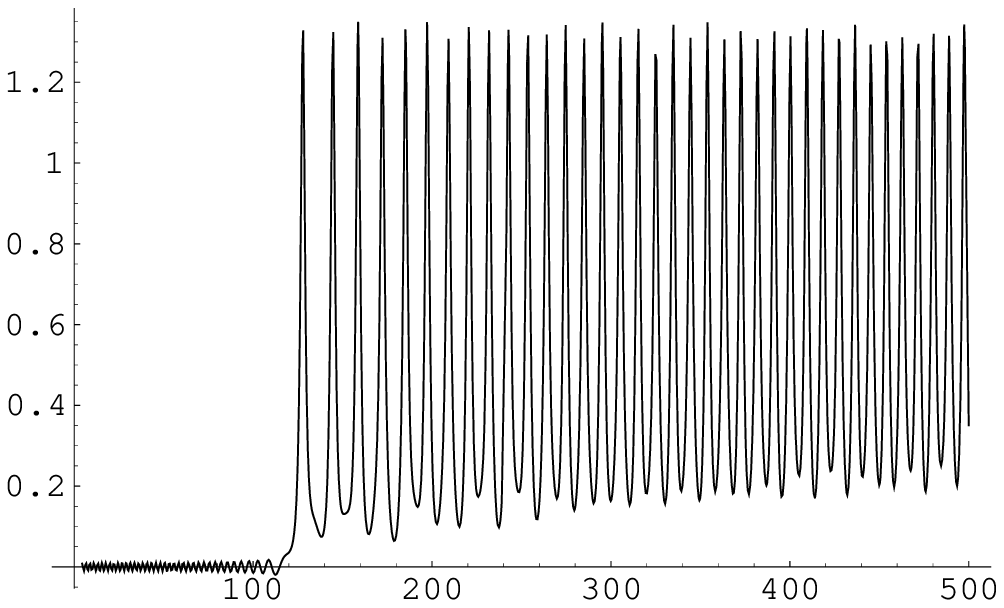}
\\
\vspace{.3cm}
\raisebox{4cm}{$\chi_2$} \hspace {-0.8cm}
\epsfxsize=2.9 in \epsfbox{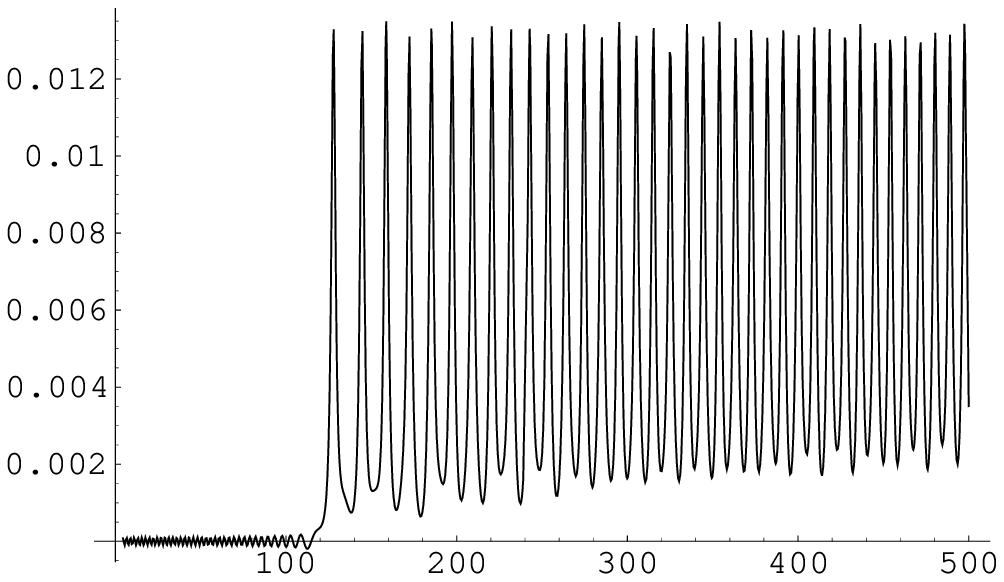}
\\
\vspace{.3cm}
\raisebox{4cm}{$H$} \hspace {-1.3cm}
\epsfxsize=2.9 in \epsfbox{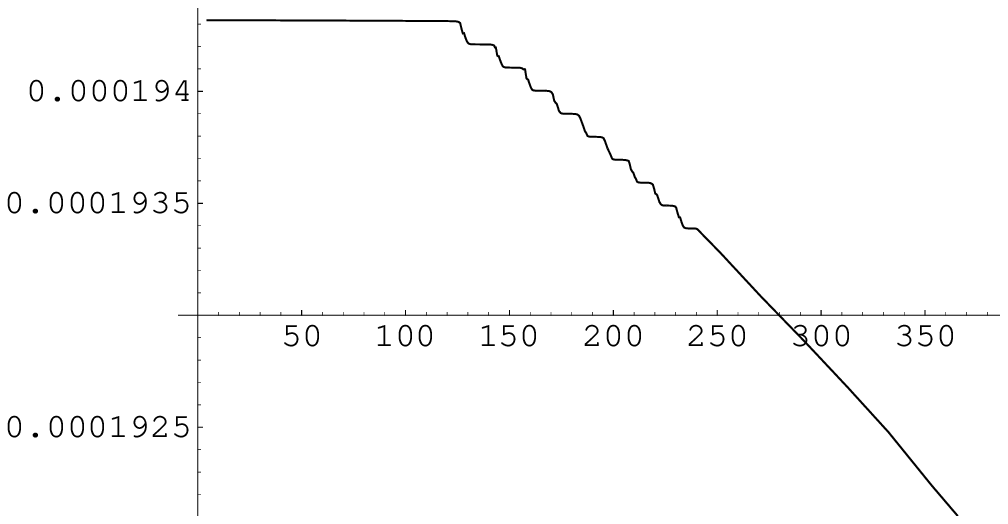}
\begin{center} $z = M \, t$ \end{center}
\vspace{.2cm}
\caption{Evolution of the background dynamics for a supersymmetric model
of hybrid inflation with $g^2 = 10^{-3}$, $M^2 / M_{pl}^2 = 10^{-12}$ 
and $M^2 / m^2 = 10^{10}$. The initial conditions are $\phi_0 = 3
M_{\rm{pl}}$, $\chi_1 = .01
M_{\rm{pl}}$, $\chi_2 = .0001 M_{\rm{pl}}$ and all the field velocities
set to zero.}
\label{fig6}  
\end{figure}

As is evident, following an initial transient period the 
three scalar fields oscillate with the same frequency. 
The results for the fluctuation variables $Q_{\phi}$, $Q_{\chi_1}$, 
$Q_{\chi_2}$ and $\zeta$ are shown in Figure 7. 
The initial perturbation amplitudes were chosen to be $Q_\phi(t_0) = 1$, 
$Q_{\chi_1} (t_0) = Q_{\chi_1} (t_0) = 10^{-4}$, and all their derivatives
set to zero for a wavelength outside the Hubble radius ($k=0$). 
As is evident, after the initial transient period, all four quantities 
grow almost with the same Floquet exponent, as expected from our
analytical analysis.

\begin{figure}
\raisebox{4cm}{$Q_\phi$} 
\epsfxsize=2.9 in \epsfbox{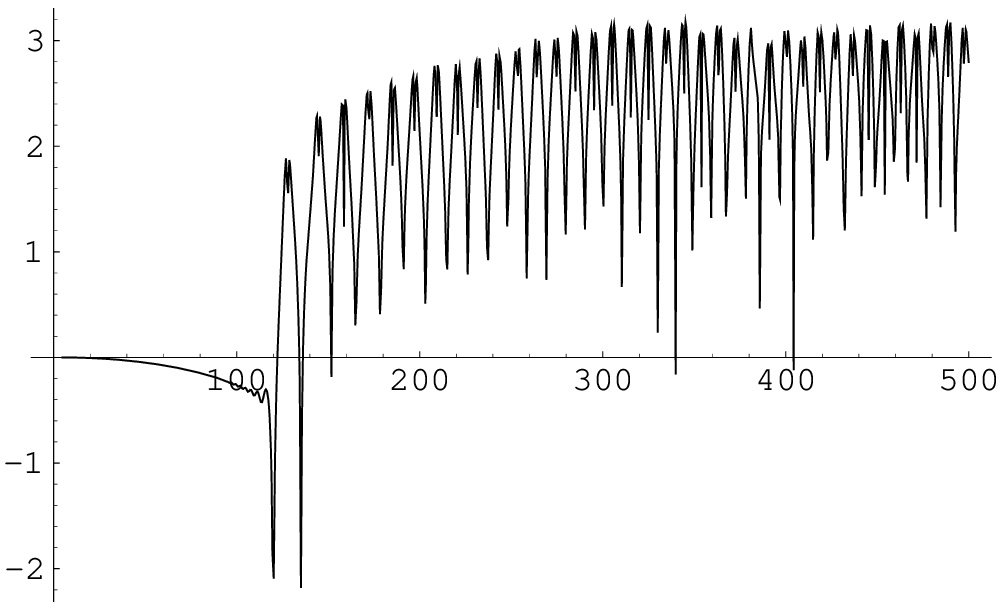}
\\
\vspace{.3cm}
\raisebox{4cm}{$Q_{\chi_1}$} \hspace {-0.4cm}
\epsfxsize=2.9 in \epsfbox{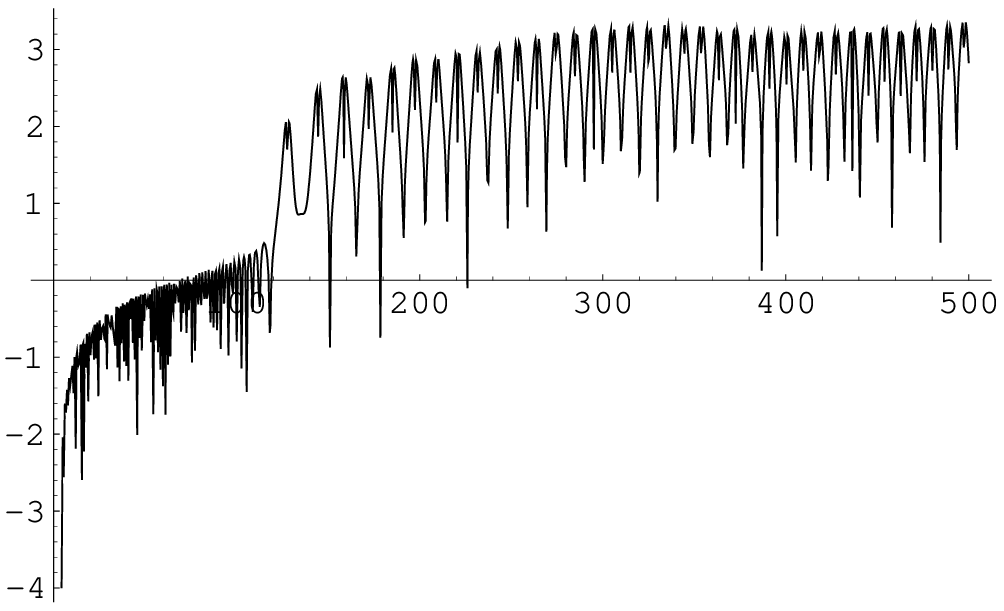}
\\
\vspace{.3cm}
\raisebox{4cm}{$Q_{\chi_2}$} \hspace {-0.4cm}
\epsfxsize=2.9 in \epsfbox{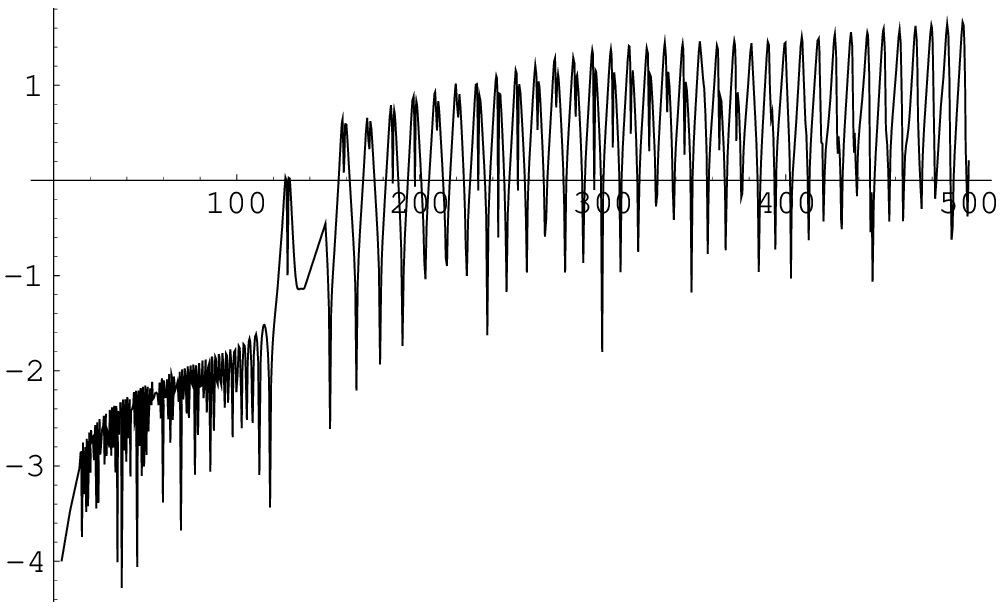}
\\
\vspace{.3cm}
\raisebox{4cm}{$(1+w)\zeta$} \hspace {-0.6cm}
\epsfxsize=2.9 in \epsfbox{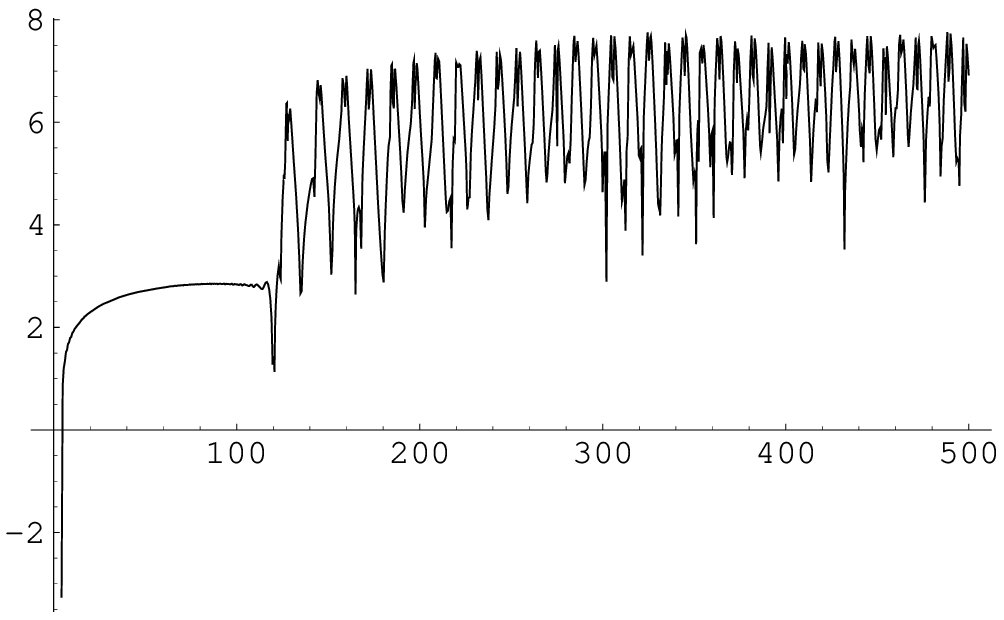}
\begin{center} $z = M \, t$ \end{center}
\vspace{.2cm}
\caption{Evolution in logarithmic scale for the fluctuation variables
$Q_{\phi}$, $Q_{\chi_1}$,
$Q_{\chi_2}$ and $(1+w)\zeta$.
The initial perturbation amplitudes were chosen to be $Q_\phi(t_0) = 1$,
$Q_{\chi_1} (t_0) = Q_{\chi_1} (t_0) = 10^{-4} M_{\rm{pl}}$, and all
their derivatives set to zero for a wavelength outside the Hubble radius
($k=0$). }
\label{fig7}
\end{figure}

At this point, an obvious question is whether the field $\chi_2$ is 
essential in order to obtain parametric resonance of super-Hubble-scale 
cosmological fluctuations. In fact, the field fluctuations $\delta \chi_1$ 
also will experience an effective negative coupling instability
\cite{BGKS}, and therefore the presence of $\chi_2$ is not essential for
this supersymmetric choice of the parameters. 
The equation of motion for $\delta \chi_1$ for a hybrid model with two
field is
\begin{equation} \label{chi1fluct}
{\ddot \delta \chi_1} + 3 H {\dot \delta \chi_1} \, 
= \, - \bigl( {{k^2} \over {a^2}} + g^2 \phi^2 
+ 3 \lambda \chi_1^2 - M^2 \bigr) \delta \chi_1 \, .
\end{equation}
The effective squared mass is large and positive during slow-rolling. At
the beginning of the transient period (when $\chi_1$ starts rolling down 
its potential but is not yet close to the minimum of the potential) 
the effective squared mass turns negative. Once $\chi_1$ gets close to its 
equilibrium position, the effective squared mass will again be large and 
positive (the factor $3$ in the third term on the r.h.s. of 
(\ref{chi1fluct}) is crucial). However, since $\chi_1$ is oscillating 
with a large amplitude, the effect of the large mass will be periodically 
cancelled out by these oscillations. 
Neglecting the expansion of the background, Eq. (\ref{chi1fluct}) 
can be written as
\begin{equation}
\chi_{1 k}^{''} + \chi_{1 k} \bigl( {{k^2} \over {2 a^2 M^2}} 
+ {{g^2 \phi^2} \over {2 M^2}} + 1 + 3 f(z) + {3 \over 2} f^2 \bigr) \, 
= \, 0 \, .
\end{equation}
{F}rom our numerical results (Figure 6) we expect that the amplitude 
of $f(z)$ will be only slightly smaller than 1. 
Hence, we expect negative coupling instability for long wavelength 
metric perturbations also in the two field case \cite{BGKS}. However, as
demonstrated below for values of the coupling constants which do not 
correspond to the supersymmetric point, the Floquet index in the two 
field case will be smaller than in the three field model.

Other interesting effects happen if we go away from the
supersymmetric point $g^2 = 2 \lambda$. 
In analogy with the results of \cite{ZMCB} which show that 
random noise in the inflaton leads to an increase in the strength of 
the parametric instability, we expect that the chaotic background dynamics
will not eliminate but rather strengthen the resonance. 
Chaotic background dynamics is expected 
for $g^2 \sim \lambda$ in hybrid models \cite{GBL}.
This effect is shown by our 
numerical simulations (Figure 8 and Figure 9) which show that the
parametric resonance of super-Hubble-scale gravitational
fluctuations for the choice $g^2 = \lambda$ is larger than in the
supersymmetric case, where no chaoticity is present \cite{BGKS}.
Figure 10 shows how the presence of the ``Goldstone" mode
$\chi_2$ changes the development of the resonance in this chaotic case.
In the two field case the Floquet index with which $\zeta$ grows is 
smaller than the corresponding index in the three field case. 

\begin{figure}
\raisebox{4cm}{$\phi$} 
\epsfxsize=2.9 in \epsfbox{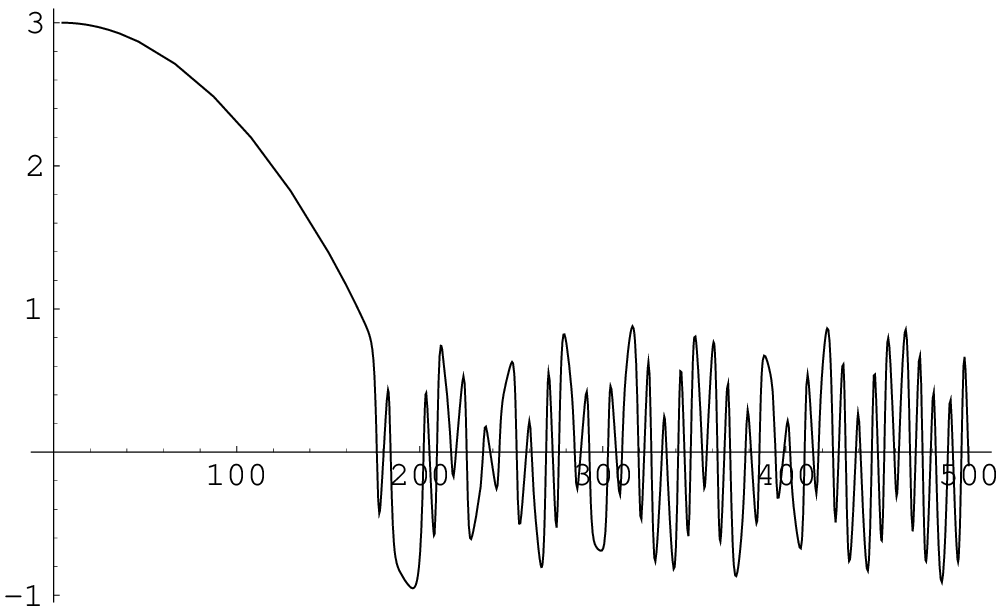}
\\
\vspace{.3cm}
\raisebox{4.1cm}{$\chi_1$} \hspace {-0.5cm}
\epsfxsize=2.9 in \epsfbox{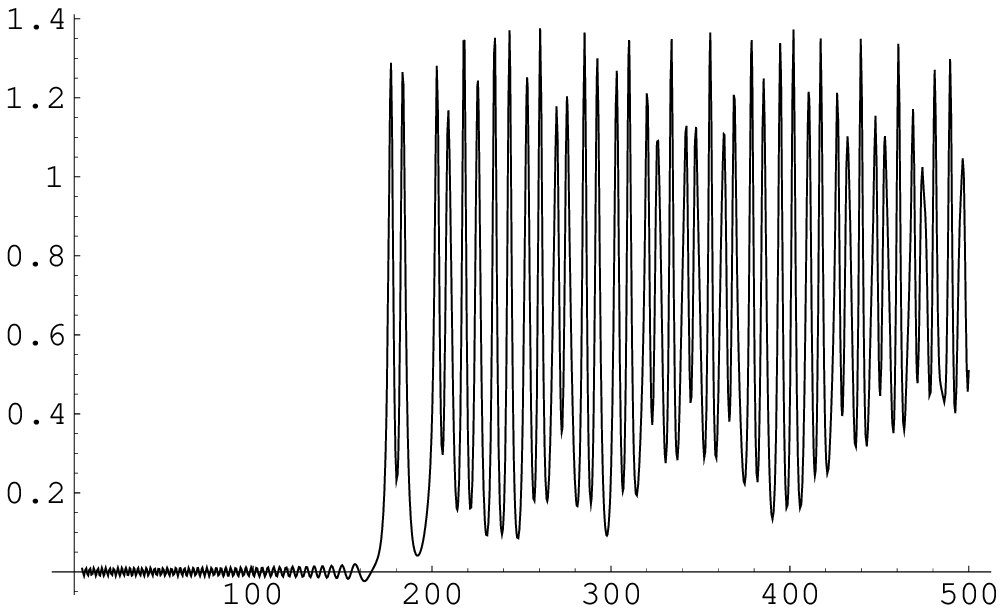}
\\
\vspace{.3cm}
\raisebox{4cm}{$\chi_2$} \hspace {-0.7cm}
\epsfxsize=2.9 in \epsfbox{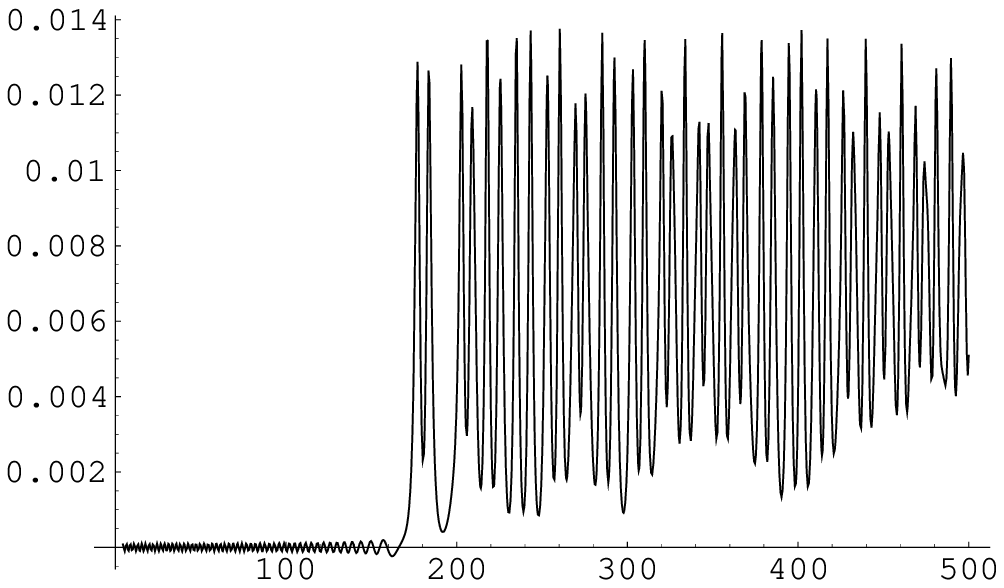}
\\
\vspace{.3cm}
\raisebox{4cm}{$H$} \hspace {-1.2cm} 
\epsfxsize=2.9 in \epsfbox{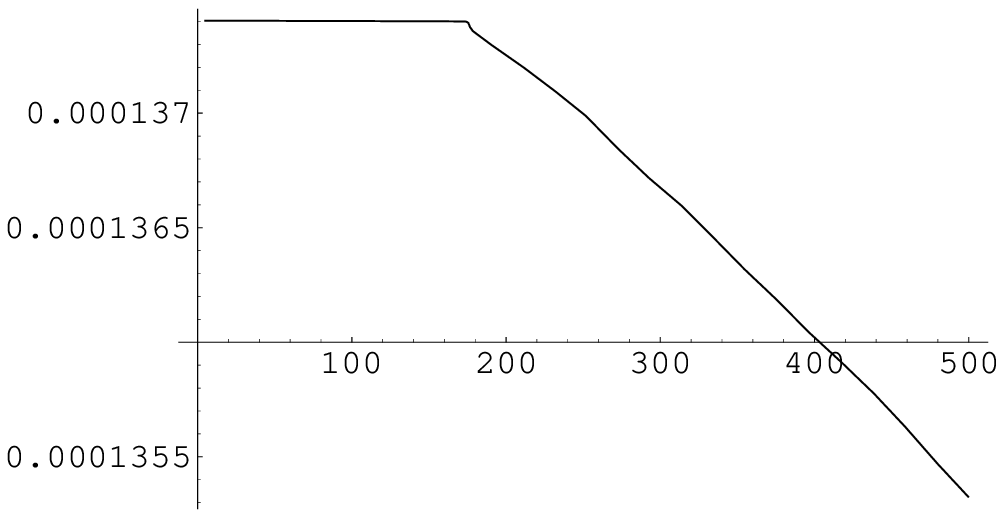}
\begin{center} $z = M \, t$ \end{center}
\vspace{.2cm}
\caption{Evolution of the background dynamics for a model of hybrid
inflation with $g^2=\lambda$. The parameters and the initial conditions
for the fields are the same of Fig. 6.}
\label{fig8}
\end{figure}

\begin{figure}
\raisebox{4cm}{$Q_\phi$} 
\epsfxsize=2.9 in \epsfbox{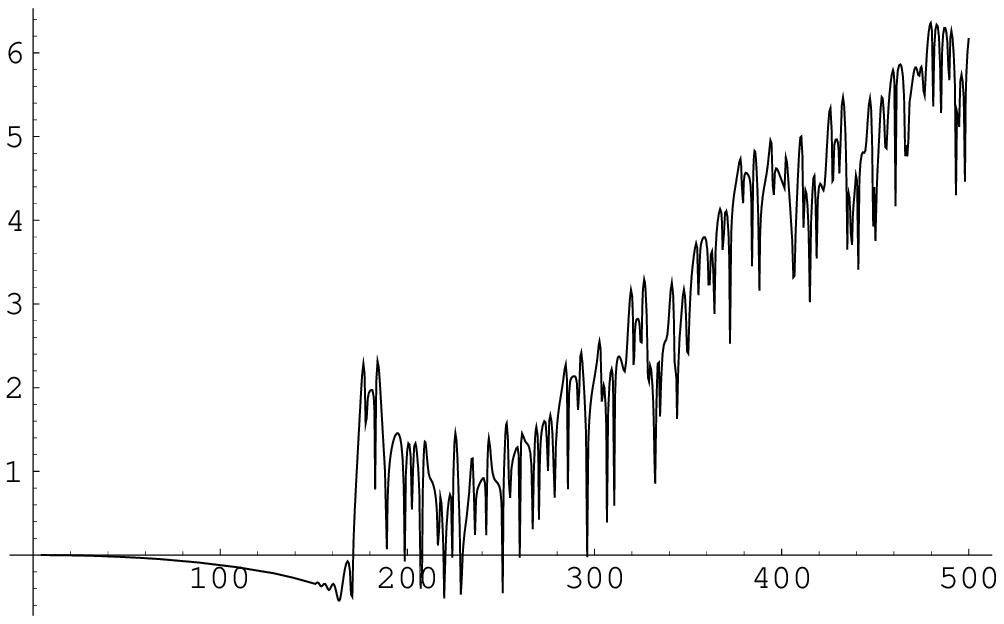}
\\
\vspace{.3cm}
\raisebox{4cm}{$Q_{\chi_1}$} \hspace {-0.5cm}
\epsfxsize=2.9 in \epsfbox{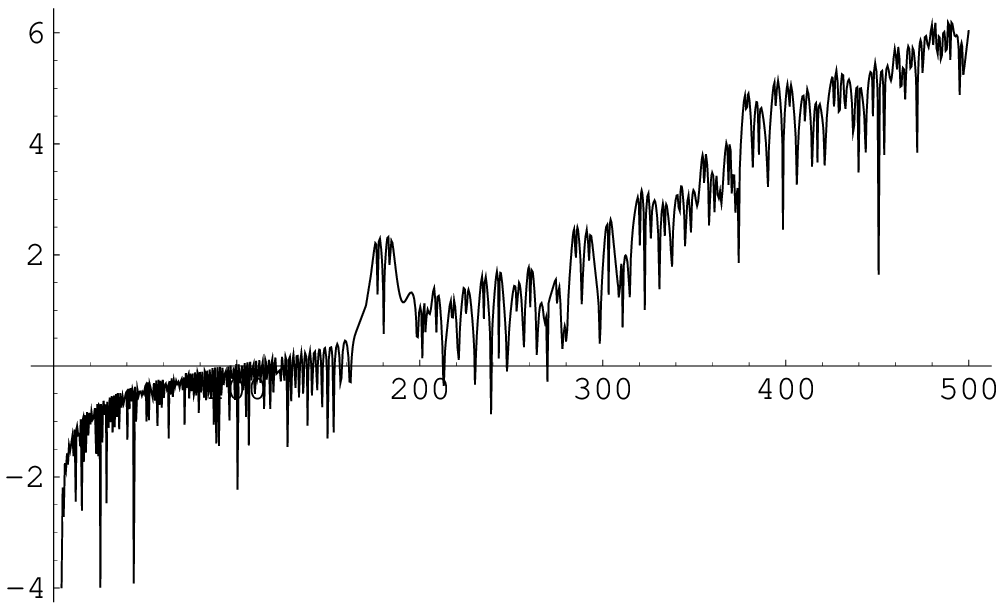}
\\
\vspace{.3cm}
\raisebox{4cm}{$Q_{\chi_2}$} \hspace {-0.5cm}
\epsfxsize=2.9 in \epsfbox{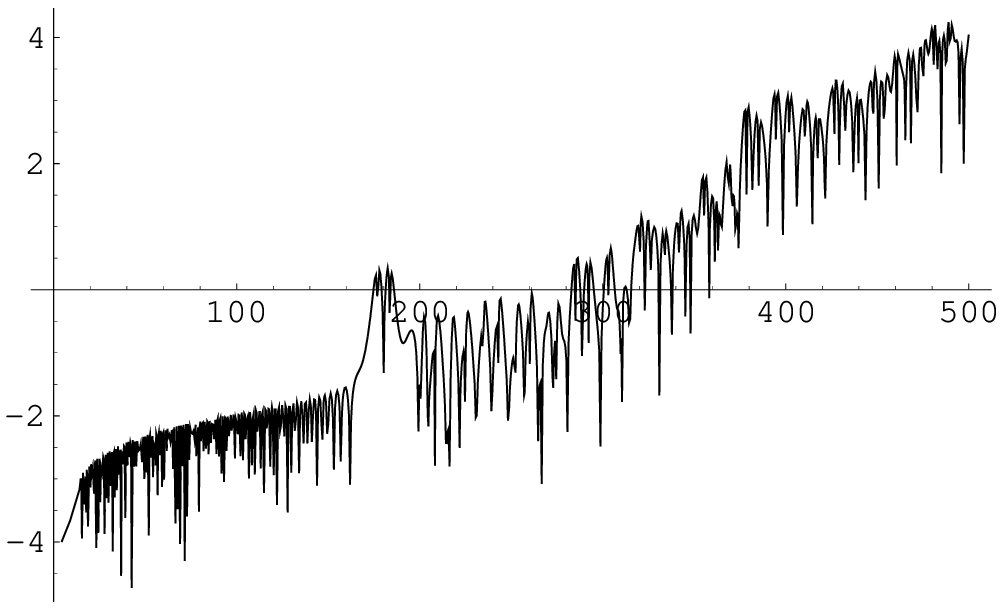}
\\
\vspace{.3cm}
\raisebox{4.2cm}{$(1+w)\zeta$} \hspace {-0.7cm}  
\epsfxsize=2.9 in \epsfbox{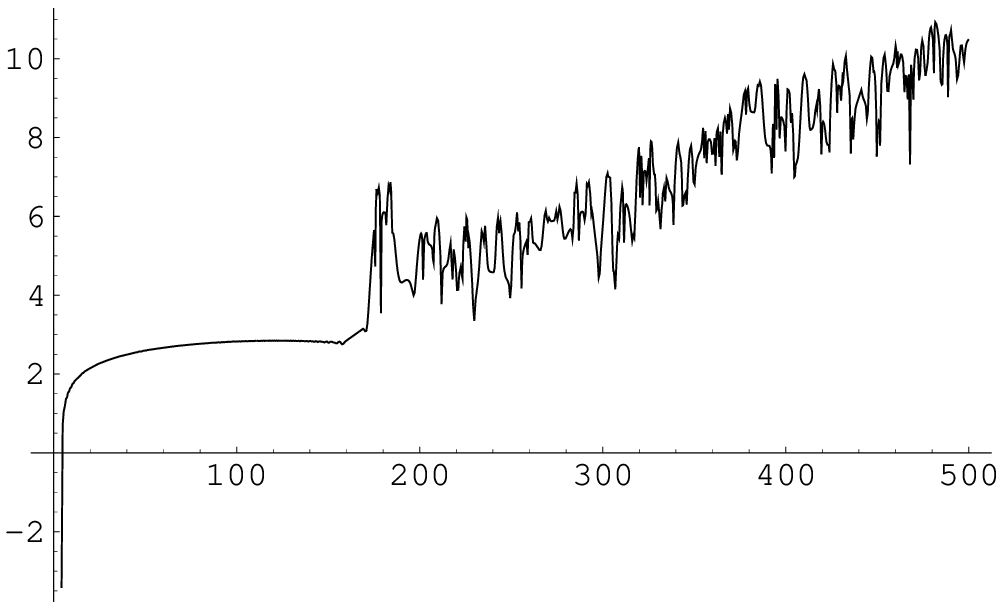}
\begin{center} $z = M \, t$ \end{center}
\vspace{.2cm}
\caption{Evolution in logarithmic scale for the fluctuation variables
$Q_{\phi}$, $Q_{\chi_1}$,
$Q_{\chi_2}$ and $(1+w)\zeta$ for the case $g^2=\lambda$ of Fig. 8.
The initial perturbation amplitudes are the same of Fig. 7.}
\label{fig9}
\end{figure}

\begin{figure}
\raisebox{4cm}{$(1+w) \zeta$} \hspace {-0.2cm}
\epsfxsize=2.9 in \epsfbox{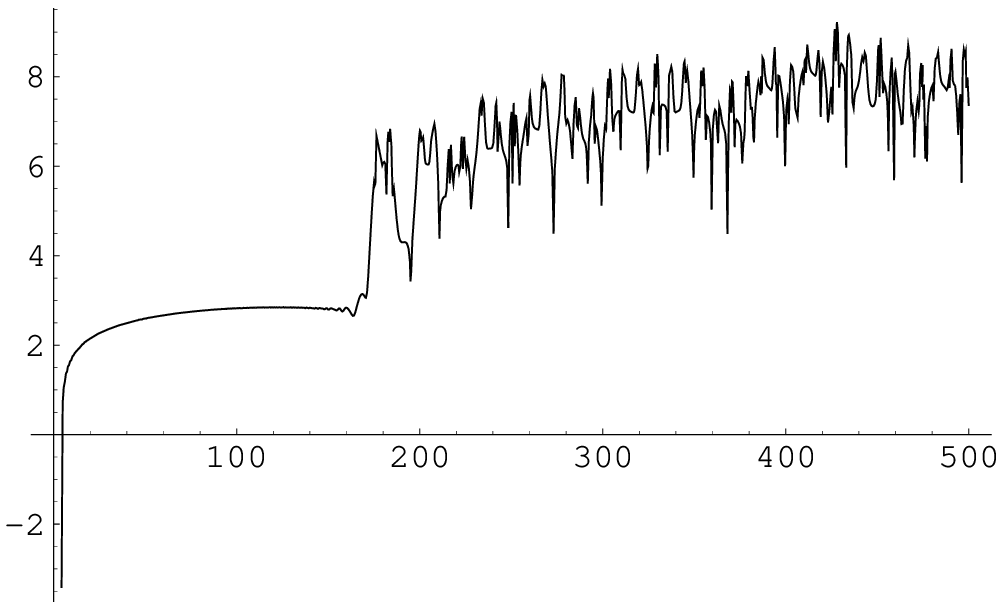}
\begin{center} $z = M \, t$ \end{center}
\vspace{.2cm}
\caption{Evolution in logarithmic scale of $(1+w)\zeta$ for the choice
$g^2=\lambda$ for a two 
field hybrid model ($\chi_2 = \delta \chi_2 =0$). The parameters, initial
conditions for the background and for the perturbation amplitudes are the 
same of Figs. 8 and 9.}
\label{fig10}
\end{figure}

\section{Discussion}

We have studied the parametric amplification of long wavelength gravitational 
fluctuations during reheating in two field inflationary Universe models. 
We have partially confirmed the results of Bassett and Viniegra
\cite{BV99} and shown that this effect is possible for certain models. 
We have established criteria under which an exponential increase 
in the amplitude of cosmological perturbations during the period 
when the inflaton field oscillates should be expected. 
It is crucial that there must be either broad-band parametric instability 
or negative coupling instability in the matter sector of the theory 
(i.e., in the absence of gravitational perturbations). This will excite 
isocurvature fluctuations during reheating. It is important that there be 
no net exponential damping in the amplitude of the isocurvature fluctuations 
before reheating. The resonance in the matter sector then induces a resonance 
in the gravitational sector provided that the background values of the
matter fields do not vanish. 
Since large coupling constants are not necessary in order to have
efficient resonance, the effect is stable
against perturbative coupling constant renormalizations. 
We have shown that in this case the resulting 
increase in the amplitude of the adiabatic mode, conveniently tracked in 
terms of the variable $\zeta$, and of isocurvature fluctuations,
tracked in terms of the non adiabatic pressure 
$p \Gamma$, is exponential and is due to the oscillations 
in the inflaton field. This means that the effect is absent if the phase 
transition is modelled by a monotonic change in $w = p / \rho$.

We then argue that the conditions under which parametric amplification 
of long wavelength gravitational fluctuations occurs are naturally 
satisfied in a class of models of hybrid inflation.  
The presence of a complex matter scalar field enhances the resonance, 
since it ensures the existence of a field which is massless in the 
true vacuum of the theory, but it is not crucial if there 
is negative coupling instability. 
However, note that the existence of massless modes is helpful for the
effect to occur. Such massless modes arise quite generically in string theory 
(see e.g. \cite{Dine} for a recent review). 
Thus, the parametric amplification of long wavelength fluctuations 
may be also present in models of inflation based on string theory.

\centerline{\bf Acknowledgements}

We are grateful to Bruce Bassett, Serguei Khlebnikov, Lev Kofman and Bill Unruh for stimulating discussions, and Jim Zibin for comments on the draft. 
R.B. wishes to thank Bill Unruh for hospitality at the 
University of British Columbia where this work was completed.
F. F. wishes to thank Brown University for hospitality. 
The research was supported in part (at Purdue) by the U.S. 
Department of Energy under Contract DE-FG02-91ER40681, TASK B, and 
(at Brown) by DE-FG02-91ER40688, TASK A.

\end{document}